\documentclass[11pt,a4paper]{article}
\usepackage[T1]{fontenc}
\usepackage[utf8]{inputenc}
\usepackage{authblk}
\usepackage{amsfonts,amsmath,amssymb} 
\usepackage{subfig}
\usepackage[ruled, noend, noline]{algorithm2e}
\usepackage{booktabs}
\usepackage{multirow}
\usepackage{acronym}
\usepackage{url}

\newcommand{\sysacro}{LEDAsig}
\newcommand{\sysdesc}{Low-dEnsity generator matrix coDe-bAsed digital signature algorithm}

\newcommand{\qcS}{\widetilde{S}}
\newcommand{\qcLam}{\widetilde{\Lambda}}
\newcommand{\qcPhi}{\widetilde{\Phi}}
\newcommand{\qcPsi}{\widetilde{\Psi}}
\newcommand{\qcPi}{\widetilde{\Pi}}
\newcommand{\qcQ}{\widetilde{Q}}
\newcommand{\qcM}{\widetilde{M}}
\newcommand{\qcR}{\widetilde{R}}

\newcommand{\qcH}{\widetilde{H}}
\newcommand{\qcG}{\widetilde{G}}
\newcommand{\qcV}{\widetilde{V}}
\acrodef{LDPC}{low-density parity-check}
\acrodef{LDGM}{low-density generator matrix}
\acrodef{MDPC}{moderate-density parity-check}
\acrodef{QC}{quasi-cyclic}
\acrodef{QC-LDPC}{quasi-cyclic low-density parity-check}
\acrodef{QC-LDGM}{quasi-cyclic low-density generator matrix}
\acrodef{QC-MDPC}{quasi-cyclic moderate-density parity-check}
\acrodef{RSA}{Rivest-Shamir-Adleman}
\acrodef{BF}{bit flipping}
\acrodef{SPA}{sum product algorithm}
\acrodef{RDF}{random difference families}
\acrodef{ISD}{information set decoding}
\acrodef{KRA}{key recovery attack}
\acrodefplural{KRA}[KRAs]{key recovery attacks}
\acrodef{DA}{decoding attack}
\acrodefplural{DA}[DAs]{decoding attacks}
\acrodef{WF}{work factor}
\acrodef{BER}{bit error rate}
\acrodef{CER}{codeword error rate}
\acrodef{BSC}{binary symmetric channel}
\acrodef{BPSK}{binary phase shift keying}
\acrodef{$2$-PAM}{binary pulse amplitude modulation}
\acrodef{AWGN}{additive white Gaussian noise}
\acrodef{LLR}{log likelihood ratio}
\acrodefplural{LLR}[LLRs]{log likelihood ratios}
\acrodef{SPA}{sum-product algorithm}
\acrodef{DFR}{decoding failure rate}
\acrodef{SL}{security level}
\acrodefplural{SL}[SLs]{security levels}
\acrodef{ECC}{elliptic curve cryptography}
\acrodef{QD}{quasi-dyadic}
\acrodef{GRS}{generalized Reed-Solomon}
\acrodef{DSA}{Digital Signature Algorithm}
\acrodef{ECDSA}{Elliptic Curve Digital Signature Algorithm}
\acrodef{KEM}{key encapsulation module}
\acrodef{PKC}{public-key cryptosystem}
\acrodef{SK}{secret key}
\acrodef{PK}{public key}
\acrodef{IND-CCA}{indistinguishability under chosen ciphertext attack}
\acrodef{IND-CCA2}{indistinguishability under adaptive chosen ciphertext attack}
\acrodef{IND-CPA}{indistinguishability under chosen plaintext attack}
\acrodef{KI}{Kobara-Imai}
\acrodef{PFS}{perfect forward secrecy}
\acrodef{NP}{nondeterministic-polynomial}
\acrodef{DRBG}{deterministic random bit generator}
\acrodef{TRNG}{true random number generator}
\acrodef{KDF}{key derivation function}
\acrodef{AKE}{authenticated key exchange}
\acrodef{SDP}{syndrome decoding problem}

\title{Design and Implementation of a Digital Signature Scheme Based on Low-density Generator Matrix Codes}
\author[1]{Marco Baldi\thanks{m.baldi@univpm.it}}
\author[2]{Alessandro Barenghi\thanks{alessandro.barenghi@polimi.it}}
\author[1]{Franco Chiaraluce\thanks{f.chiaraluce@univpm.it}}
\author[2]{Gerardo Pelosi\thanks{gerardo.pelosi@polimi.it}}
\author[3]{Joachim Rosenthal\thanks{rosenthal@math.uzh.ch}}
\author[1]{Paolo Santini\thanks{p.santini@pm.univpm.it}}
\author[3]{Davide Schipani\thanks{davide.schipani@math.uzh.ch}}
\affil[1]{Universit\`a Politecnica delle Marche, Ancona, Italy}
\affil[2]{Politecnico di Milano, Milano, Italy}
\affil[3]{University of Zurich, Zurich, Switzerland}
\date{}

\begin{document}
\maketitle
	
\begin{abstract}
In this paper we consider a post-quantum digital signature scheme based on
low-density generator matrix codes and propose efficient algorithmic solutions
for its implementation.
We also review all known attacks against this scheme and derive closed-form 
estimates of their complexity when running over both classical and quantum
computers. Based on these estimates, we propose new parametrization for the
considered system to achieve given pre-quantum and post-quantum security levels.
Finally, we provide and discuss performance benchmarks obtained through a 
suitably developed and publicly available reference implementation of the 
considered system.
\end{abstract}
%
%
\section{Introduction}\label{sec:intro}
Devising efficient post-quantum cryptographic schemes is a primary challenge, as
also witnessed by the recently started NIST post-quantum standardization initiative \cite{NISTcall2016}.
Among post-quantum cryptographic primitives, solutions based on error correcting
codes and lattices play a primary role.

In this paper we deal with post-quantum cryptographic primitives based on codes
and, in particular, code-based digital signatures.
While it is relatively simple to devise code-based public key encryption schemes,
mostly derived from the well-known McEliece system \cite{McEliece1978}, the same 
cannot be said for digital signature schemes. In fact, code-based public-key 
encryption schemes are characterized by an expansion of the plaintext into the 
ciphertext, due to the redundancy added by encoding.
Such an expansion results in the fact that some of the bit-strings of the 
same length of a ciphertext do not belong to the encryption function codomain. 
Therefore, it is not possible to exploit the same symmetry present in, e.g., the 
\ac{RSA} scheme, to derive a signature scheme from a public key encryption
cryptosystem.

This makes the problem of finding secure yet efficient code-based digital 
signature schemes a challenging one.
Currently, the scheme introduced by Curtois, Finiasz and Sendrier (CFS) \cite{Courtois2001} is the best known solution to this problem, withstanding seventeen years of cryptanalysis.
The main drawback of this scheme is the need of decoding any syndrome vector obtained as the hash of the message to be signed, which is addressed appending a
counter to the message or performing complete decoding.
This solution however yields choices of the code parameters resulting in high
complexity \cite{Finiasz2011} and may weaken the system security 
\cite{Faugere2013}.
More recent approaches exploit different families of codes, such as \ac{LDGM} 
codes \cite{Baldi2013} and codes in the $(U | U+V)$ form 
\cite{2017arXiv170608065D},  in order to design more practical code-based digital 
signature schemes.

In this paper we focus on the former solution, and describe a code-based digital signature scheme we name \sysdesc{} (\sysacro{}).
It implements and improves the \ac{LDGM} code-based digital signature scheme 
proposed in \cite{Baldi2013c}, that is standing since 2013 as a very fast code-
based digital signature scheme with very compact public keys.
In fact, this system has been implemented on embedded hardware achieving the 
fastest implementation of code-based signatures in open literature, with a 
signature generation throughputs of around $60,000$ signatures per second 
\cite{Hu2017}.

This code-based signature scheme is characterized by very fast key generation, 
signature generation and signature verification procedures.
This is achieved by exploiting a special instance of the \ac{SDP} which allows 
to reduce decoding to a straightforward vector manipulation.
This is done by considering only a subset of all possible syndromes, formed by 
those having a fixed and low Hamming weight.
For this reason, we can say that \sysacro{} relies on the \textit{sparse} 
\ac{SDP}, which however is not easier to solve than the general \ac{SDP} without 
efficient algorithms exploiting the secret structure of the code.

The main known attacks against \sysacro{} are those already devised against the 
system in \cite{Baldi2013c}, plus statistical attacks recently introduced in 
\cite{Phesso2016}.
As shown in \cite{Phesso2016}, the digital signature scheme proposed in 
\cite{Baldi2013c} can use the same keypair to perform a limited amount of 
signatures, before this exposes the system to statistical attacks that may be 
able to recover the secret key.
\sysacro{} defines new choices of the system parameters which allow to achieve a 
reasonably long lifespan for each key pair.
Besides detailing recent statistical attacks, we carefully analyze all known 
attacks and provide a parametrization for \sysacro{} to achieve some computational 
security guarantees, taking into account the cost reduction which follows from 
the use of a quantum computer in the solution of the underlying  computationally 
hard problems.
We also provide efficient algorithmic solutions for the implementation of all the 
\sysacro{} functions.
These solutions have been included in a reference software implementation of 
\sysacro{} that is publicly available in \cite{LEDAcrypt}.
Based on this implementation, we carry out performance benchmarks of \sysacro{} 
and provide performance figures that highlight its benefits in terms of signature 
generation and verification time.

The paper is organized as follows.
In Section \ref{sec:description} we describe the scheme and the efficient 
algorithmic solutions we propose for its implementation.
In Section \ref{sec:security} we consider all known attacks that can be mounted 
against \sysacro{} and provide complexity estimates by considering both classical 
and quantum computers.
In Section \ref{sec:instances} we design some system instances to achieve given 
\acp{SL}.
In Section \ref{sec:performance} we assess performance based on the reference 
implementation of \sysacro{}.
In Section \ref{sec:conclusion} we provide some conclusive remarks.
%
%
\section{Description of the Scheme}\label{sec:description}
Following \cite{Baldi2013c}, in \sysacro{} the secret and the public keys are the 
characteristic matrices of two linear block codes: a private \ac{QC-LDGM} code and a 
public \ac{QC} code derived from the former. Some background concepts about these codes 
are recalled in Section \ref{subsec:codingbackground}.
In the description of the cryptoscheme, two public functions are used: a hash 
function $\mathcal{H}$ and a function $\mathcal{F}_\Theta$ that converts the 
output vector of $\mathcal{H}$ into a sparse vector $s$ with length $r$ and 
weight $w$ $(\ll r)$. 
The vector $s$ is a public syndrome vector resulting from the signature 
generation procedure.
The output of $\mathcal{F}_\Theta$ is uniformly distributed over all the $r$-bits
long vectors with weight $w$, and depends on a parameter $\Theta$, which is 
chosen for each message to be signed and is made public by the signer.

The design of $\mathcal{F}_\Theta$ is discussed next, where we provide a procedural description of the main steps of \sysacro{}, i.e., key generation, signature 
generation and signature verification.
We also provide some methods to accelerate the generation of the elements of 
the private key and to guarantee the non-singularity condition which is required 
for some of the involved matrices.
Efficient representations of the matrices involved in \sysacro{} are also 
introduced.
In the procedural descriptions we consider the following functions:
\begin{itemize}
\item \texttt{randGen}$(x,y)$: generates $y$ distinct integers in $\{0,1, \ldots, x-1\}$;
\item \texttt{matrGen}$(x,y)$: generates a random binary $x\times y$ matrix; 
\item \texttt{circGen}$(x,w)$: generates a random $x\times x$ circulant matrix with row (and column) weight equal to $w$;
\item \texttt{permGen}$(x)$: generates a random $x\times x$ permutation matrix. 
\end{itemize}
We use $\otimes$ to denote the Kronecker product, while the classical matrix 
product is denoted with $\cdot$ only when the Kronecker product appears in the 
same equation, otherwise it is omitted.
We denote as $A_{i,j}$ the element of matrix $A$ in the $i$-th row and 
$j$-th column.
We use $0_{n\times m}$ to denote the $n\times m$ null matrix and $1_{n\times m}$
to denote the $n\times m$ matrix with all entries equal to one.

\subsection{Coding Background\label{subsec:codingbackground}}

Let $\mathbb{F}_{2}^{k}$ denote the $k$-dimensional vector space defined over 
the binary field $\mathbb{F}_{2}$.
A binary linear block code, denoted as $\mathcal{C}\left(n,k\right)$, is defined 
as a bijective linear map $\mathcal{C}\left(n,k\right): \mathbb{F}_{2}^{k}\to 
\mathbb{F}_{2}^{n}$, $n,k\in \mathbb{N}$, $0 < k < n$, 
between any binary $k$-tuple (i.e., an information word) and a binary $n$-tuple 
(denoted as codeword).
The value $n$ is known as the length of the code, while $k$ denotes its 
dimension.
A generator matrix $G$ (resp. parity-check matrix $H$) for 
$\mathcal{C}\left(n,k\right)$ is a matrix whose row span (resp. kernel) 
coincides with the set of codewords of $\mathcal{C}\left(n,k\right)$.

A binary linear block code is said to be \ac{LDGM} if at least one of its generator matrices is sparse, i.e., has a fraction $\ll 1/2$ of its entries set to one. 
\sysacro{} uses a secret binary \ac{LDGM} code with length $n$ and dimension $k$, characterized by a generator matrix in the systematic form
\begin{equation}
G = [I_k | V],
\label{eq:GLDGM}
\end{equation}
where $I_k$ is the $k \times k$ identity matrix and $V$ is a sparse $k \times r$ 
matrix (with $r = n - k$ being the code redundancy).
As it will be shown next, a special form of $V$ is considered, known as \ac{QC} 
form, which makes the \ac{LDGM} code a \ac{QC} code as well.
The rows of $G$ have fixed Hamming weight $w_g \ll n$, which means that $V$ has constant row weight equal to $w_g-1$.

Due to their sparse nature, it is very likely that, by adding two or more rows 
of the generator matrix of an LDGM code, a vector with Hamming weight $> w_g$ is obtained.
If the linear combination of any group of rows of $G$ yields a codeword with weight greater than or equal to $w_g$, then the LDGM code has minimum distance $w_g$. 
This is even more likely if the rows of $G$ are chosen in such a way as to be 
quasi-orthogonal, that is, with a small number of overlapping ones.

The code defined by $G$ in eq. \eqref{eq:GLDGM} admits a sparse parity-check 
matrix $H$ in the form
\begin{equation}
H = [V^T | I_r],
\label{eq:sysH}
\end{equation}
where $I_r$ is the $r \times r$ identity matrix.
Due to the sparsity of $V$, the parity-check matrix $H$ in eq.~\eqref{eq:sysH} 
is a sparse matrix as well.
Therefore, an \ac{LDGM} code with generator matrix as in eq.~\eqref{eq:GLDGM} 
also is a \ac{LDPC} code.

The special class of \ac{LDGM} codes used in \sysacro{} is that of \ac{QC-LDGM} 
codes, having generator and parity-check matrices formed by circulant bocks 
with size $p \times p, p \in \{2, 3, \ldots, r\}$.
In fact, the \ac{QC} property allows to reduce the memory needed to store these 
matrices, and yields important advantages in terms of algorithmic complexity.
In case of a \ac{QC-LDGM} code, the $k \times r$ matrix $V$ in 
eq.~\eqref{eq:GLDGM} and eq.~\eqref{eq:sysH} is denoted as $\widetilde{V}$ and has 
the following general form
\begin{equation}
\widetilde{V} = \left[
\begin{array}{ccccc}
\widetilde{V}_{0,0} & \widetilde{V}_{0,1} & \widetilde{V}_{0,2} & \ldots & \widetilde{V}_{0,r_0-1} \\
\widetilde{V}_{1,0} & \widetilde{V}_{1,1} & \widetilde{V}_{1,2} & \ldots & \widetilde{V}_{1,r_0-1} \\
\widetilde{V}_{2,0} & \widetilde{V}_{2,1} & \widetilde{V}_{2,2} & \ldots & \widetilde{V}_{2,r_0-1} \\
\vdots & \vdots & \vdots & \ddots & \vdots \\
\widetilde{V}_{k_0-1,0} & \widetilde{V}_{k_0-1,1} & \widetilde{V}_{k_0-1,2} & \ldots & \widetilde{V}_{k_0-1,r_0-1} \\
\end{array}
\right],
\label{eq:CQCLDGM}
\end{equation}
where $\widetilde{V}_{i,j}$ represents either a sparse circulant matrix or a null 
matrix with size $p \times p$.
Hence, in this case the code length, dimension and redundancy are 
$n=n_0p$, $k=k_0p$ and $r=(n_0-k_0)p=r_0p$, respectively.
For the rest of the paper, we will use the superscript $\sim$ to denote \ac{QC} 
matrices and, for a given \ac{QC} matrix $\widetilde{A}$, we refer to its circulant block 
at position $(i,j)$ as $\widetilde{A}_{i,j}$.

Since a circulant matrix is defined by one of its rows (conventionally the first),
storing a binary matrix $\widetilde{V}$ as in eq. \eqref{eq:CQCLDGM} requires
$k_0 r_0 p$ bits, yielding a reduction by a factor $p$ with respect to a matrix with a 
general form.
Moreover, given the sparse form of the matrices, a further size reduction 
can be achieved storing only the positions of the set coefficients of each first
row of a circulant block.

The set of $p\times p$ binary circulant matrices form a ring under the 
operations of modulo-2 matrix addition and multiplication.
The zero element is the all-zero matrix, and the identity element is the 
$p\times p$ identity matrix.
If we consider the algebra of polynomials $\mathrm{mod}\left(x^{p}-1\right)$ over 
$\mathbb{F}_2$, $\mathbb{F}_2[x]/\langle x^p +1 \rangle$, the following map is an 
isomorphism between this algebra and that of $p\times p$ circulant matrices over 
$\mathbb{F}_2$
\begin{equation}
\widetilde{A} \leftrightarrow a\left(x\right)=\sum_{i=0}^{p-1}a_{i} x^{i}.
\label{eq:CircPoly}
\end{equation}
According to eq. \eqref{eq:CircPoly}, any binary circulant matrix is associated 
to a polynomial in the variable $x$ having coefficients over $\mathbb{F}_2$ 
which coincide with the entries in the first row of the matrix, i.e.,
\begin{equation}
a\left(x\right)=a_{0}+a_{1}x+a_{2}x^{2}+a_{3}x^{3}+\cdots+a_{p-1}x^{p-1}\label{eq:CircPoly2}.
\end{equation}
Also according to eq. \eqref{eq:CircPoly}, the all-zero circulant matrix 
corresponds to the null polynomial and the identity matrix to the unitary 
polynomial.
In the same way, the set of $r_0p \times r_0p$ \ac{QC} matrices formed by 
circulant blocks of size $p \times p$ is a ring under the standard operations of 
modulo-2 matrix addition and multiplication. 
The null element corresponds to the null $r_0p \times r_0p$ matrix, the identity element is the $r_0p\times r_0p$ identity matrix $I_{r_0p}$.
Matrices in \ac{QC} form can be efficiently represented by the polynomials associated to the circulant blocks, leading to very compact representations.

The \ac{LDGM} codes used in \sysacro{} are described by generator matrices with 
constant row weight $w_g\ll n$, a feature which ie employed to easily 
obtain a random codeword $c$ with weight $w_c \approx m_g w_g$, with $m_g$ 
being a small integer.
In fact, since the rows of the generator matrix are sparse, it is very likely 
that, by adding together a few of them, the Hamming weight of the resulting 
vector is about the sum of the Hamming weights of its addends, bar some cancellations
due to overlapping ones.
If the sum of a set of rows does not fit the desired weight $w_c$, some other 
row can be added, or some row replaced, or another combination of rows can be 
tested, in order to approach $w_c$.
In fact, using codewords with weight slightly smaller than $w_c$ may still allow 
achieving the target security level.
In any case, generating a codeword with weight equal or almost equal to $w_c$ 
can be accomplished very quickly.

Based on these considerations, the number of random codewords with weight close 
to $w_c$ which can be easily generated at random from an \ac{LDGM} code having 
row weight of $G$ equal to $w_g$, with $w_g$ dividing $w_c$, can be roughly 
estimated as
\begin{equation}
A_{w_c} \approx \binom{k}{\frac{w_c}{w_g}}.
\end{equation}

\subsection{Private Key Generation}
\label{sec:KeyGen}
The private key in \sysacro{} includes the characteristic matrices of an 
\ac{LDGM} code $\mathcal{C}$ with length $n$, dimension $k$ and co-dimension 
$r=n-k$.
In particular, we consider circulant matrices of size $p$, and so we have 
$n_0=n/p$, $k_0=k/p$ and $r_0=r/p$.
We denote the generator and the parity-check matrix as $\widetilde{G}$ and 
$\widetilde{H}$, respectively.
Because of their systematic forms \eqref{eq:GLDGM} and \eqref{eq:sysH}, these
matrices can be represented just through $\widetilde{V}$.
The public key is a dense $r_0p \times n_0p$ \ac{QC} matrix.

\subsubsection{Generation of $\widetilde{H}$ and $\widetilde{G}$}
The generator and parity-check matrices of the secret \ac{LDGM} code have the 
form \eqref{eq:GLDGM} and \eqref{eq:sysH}, respectively.
They are both obtained starting from $\widetilde{V}$, which is generated as 
described in Algorithm \ref{alg:genV}.

\begin{algorithm}[!t]
{\small 
\LinesNumbered
\DontPrintSemicolon
\caption{Generation of $\widetilde{V}$\label{alg:genV}}
\KwIn{
      $p$: size of a circulant block \newline
      $r_0$: code dimension divided by circulant block size \newline 
      $k_0$: code redundancy divided by circulant block size \newline
      $w_g$: weight of a row of $\widetilde{V}$}
\KwOut{$\widetilde{V}$}
\BlankLine
$\widetilde{V}\leftarrow 0_{k_0p \times r_0p}$\;
\For{$i\leftarrow 0$ $\mathbf{to}$ $k_0-1$}{
$[t_0,t_1,\ldots,t_{w_g-2}]\leftarrow $\texttt{randGen}$(r_0, w_g-1)$\;
\For{$j\leftarrow 0$ $\mathbf{to}$ $w_g-2$}{
       $\widetilde{V}_{i,t_j}\leftarrow$ \textsc{circGen}$(p,1)$\;
}
}
\KwRet$\widetilde{V}$
}
\end{algorithm}

\subsubsection{Generation of $\widetilde{S}$ and $\widetilde{S}^{-1}$}

The matrix $\widetilde{S}$ is a $pn_0\times pn_0$ binary matrix, with constant 
row and column weight equal to $m_s$.
There are several methods for generating such a matrix.
We consider the procedure described in Algorithm \ref{alg:genS}, which allows an efficient computation of $\widetilde{S}^{-1}$.

\begin{algorithm}[!t]
{\small 
\LinesNumbered
\DontPrintSemicolon
\caption{Generation of $\widetilde{S}$\label{alg:genS}}
\KwIn{ $p$: size of a circulant block \newline
       $n_0$:code size divided by circulant block size \newline
       $m_s$: weight of a row of $\widetilde{S}$}
\KwOut{$\widetilde{S}$}
$\widetilde{S}\leftarrow 0_{n_0\times n_0}$\;
$\widetilde{E}\leftarrow $\texttt{circGen$(n_0,m_s)$}\;
$\Pi^{(1)}\leftarrow$\texttt{permGen$(n_0)$}\;
$\Pi^{(2)}\leftarrow$\texttt{permGen$(n_0)$}\;
$E'\leftarrow \Pi^{(1)}\cdot \widetilde{E} \cdot \Pi^{(2)}$\;
$\widetilde{\Lambda}\leftarrow 0_{p \times n_0p }$\;
$\widetilde{\Phi}\leftarrow 0_{p\times n_0p }$\;
\For{$i\leftarrow 0$ $\mathbf{to}$ $n_0-1$}{
   $\widetilde{\Lambda}_{i}\leftarrow$ \textsc{circGen}$(p,1)$\;
   $\widetilde{\Phi}_{i}\leftarrow$ \textsc{circGen}$(p,1)$\;   
}
\For{$i\leftarrow 0$ $\mathbf{to}$ $n_0-1$}{
   \For{$j\leftarrow 0$ $\mathbf{to}$ $n_0-1$}{
	 \If{$E'_{i,j}=1$}{
	   $\widetilde{S}_{i,j}\leftarrow \widetilde{\Lambda}_i\cdot \widetilde{\Phi}_j$\;
	 }
     \Else{
           $\widetilde{S}_{i,j}\leftarrow 0_{p\times p}$\;
     }
   }
}
}
\end{algorithm}

According to Algorithm \ref{alg:genS}, and denoting as $Diag\left(A_0,A_1,\cdots,A_{l-1}\right)$ a $(pl) \times (pl)$ diagonal matrix with $p \times p$ 
blocks $A_0,A_1,\cdots,A_{l-1}$ along the main diagonal, we can write:
\begin{align}
\label{eq:Skron}
\qcS \nonumber & = Diag\left(\qcLam_0,\qcLam_1,\cdots,\qcLam_{n_0-1}\right) \left[\left( \Pi^{(1)} \widetilde{E}\cdot \Pi^{(2)}\right) \otimes I_p\right]\cdot Diag\left(\qcPhi_0,\qcPhi_1,\cdots,\qcPhi_{n_0-1}\right) \\
& = Diag\left(\qcLam_0,\qcLam_1,\cdots,\qcLam_{n_0-1}\right)\cdot \left( \Pi^{(1)}\otimes I_p\right)\cdot \left(\widetilde{E}\otimes I_p\right)\cdot \left( \Pi^{(2)}\otimes I_p\right)\cdot Diag\left(\qcPhi_0,\qcPhi_1,\cdots,\qcPhi_{n_0-1}\right)\\\nonumber
& = \qcS \qcPi_{\Lambda} \cdot (\widetilde{E}\otimes I_p) \cdot .
\end{align}
where
\begin{equation}
\begin{array}{r@{}l}
\qcPi_{\Lambda} & = Diag\left(\qcLam_0,\qcLam_1,\cdots,\qcLam_{n_0-1}\right)\cdot \left( \Pi^{(1)}\otimes I_p\right),\\
\qcPi_{\Phi} & = \left( \Pi^{(2)}\otimes I_p\right)\cdot Diag\left(\qcPhi_0,\qcPhi_1,\cdots,\qcPhi_{n_0-1}\right).
\end{array}
\end{equation}
Based on eq. \eqref{eq:Skron}, we have
\begin{equation}
\qcS^{-1}=\qcPi_{\Phi}^{-1} \cdot (\widetilde{E}^{-1}\otimes I_p) \cdot \qcPi_{\Lambda}^{-1}.
\end{equation}
Now, since $\qcPi_{\Lambda}$ and $\qcPi_{\Phi}$ are permutation matrices, their 
inverses correspond to their transposes, yielding
\begin{equation}
\qcS^{-1}=\qcPi_{\Phi}^{T} \cdot (\widetilde{E}^{-1}\otimes I_p) \cdot \qcPi_{\Lambda}^{T}.
\end{equation}

This approach allows to achieve significant speedups in the inversion of $\qcS$, 
since the most complex part of the computation is the inversion of $\widetilde{E}$, 
which is an $n_0\times n_0$ matrix, with $n_0$ being typically two orders of 
magnitude smaller than the code length $n$.
The existence of $\widetilde{E}^{-1}$ is sufficient to guarantee that $\qcS$ is 
non singular.
If we choose $m_s$ odd and $n_0$ such that $(x^{n_0}+1)/(x+1)\in\mathbb{F}_2[x]$ is 
irreducible \cite{Xing:2003:CTF:861290}, then $\widetilde{E}^{-1}$ always exists. 

\subsubsection{Generation of $\qcQ$ and $\qcQ^{-1}$ \label{subsubsec:qcQ}}

The matrix $\qcQ$ is a $pr_0\times pr_0$ matrix obtained as
\begin{equation}
\qcQ = \qcR + \qcM,
\label{eq:qcQ}
\end{equation}
where $\qcR$ is a dense matrix with rank $z\ll pr_0$ and $\qcM$ is a sparse 
matrix. 
The density of $\qcM$ can be considered as a parameter of the system design.
In the following, we assume that $\widetilde{M}$ has constant row and column 
weight equal to $1$ (i.e., it is a permutation matrix), since this choice 
has several advantages from the complexity standpoint. 
In particular, we propose the following construction for the matrices in the 
r.h.s. of \eqref{eq:qcQ}:
\begin{equation}
\begin{cases}
\qcR=\left(A\cdot B^T \right)\otimes 1_{p\times p}=\left(A\otimes 1_{p\times 1}\right)\cdot(B^T\otimes 1_{1\times p}),\\
\qcM=(\Pi\otimes I_p)\cdot \qcPsi,
\end{cases}
\label{eq:RM}
\end{equation}
in which $A$ and $B$ are two $r_0 \times z$ random binary matrices, 
$\qcPsi = Diag(\qcPsi_0,\qcPsi_1,\cdots,\qcPsi_{r_0-1})$ denotes a \ac{QC} diagonal 
matrix having the circulant permutation matrices $\qcPsi_i$ along the main 
diagonal and $\Pi$ is an $r_0\times r_0$ permutation matrix.
We choose $z<r_0$, such that the matrix $A\cdot B^T$ has maximum rank $z<r_0$; 
since $1_{p\times p}$ has rank equal to $1$, the rank of $\qcR$ equals the one 
of $A\cdot B^T$, and so cannot be larger than $z$.
The overall row and column weight of $M$ will be denoted as $m_T$ in the following. 
As we show next, the inverse of $\qcQ$ can be easily computed and its existence depends on the choice of $\qcM$ and $\qcR$.
This is already considered in Algorithm \ref{alg:genQ} for their generation.

\begin{algorithm}[!t]
{\small 
\LinesNumbered
\DontPrintSemicolon
\caption{Generation of $\qcM, A, B^T$\label{alg:genQ}}
\KwIn{ $p$: size of a circulant block \newline
       $r_0$: code redundancy divided by circulant block size \newline
       $z$: maximum rank of $AB^T$}
\KwOut{$\qcM, A, B^T, D$}
\BlankLine
$D\leftarrow 0_{r_0\times r_0}$\;
\While{$\mathrm{det}(D) = 0$ \textbf{and} $p \bmod 2 = 1$}{
   $\Pi\leftarrow$\textsc{permGen}$(r_0)$\;
   $A\leftarrow$\textsc{matrGen}$(z,r_0)$\;
   $B^T\leftarrow$\textsc{matrGen}$(r_0,z)$\;
   $D\leftarrow I_z+B^T\Pi^T A$\;
 }
$\qcM\leftarrow 0_{r_0p\times r_0p}$\;	
\For{$i\leftarrow 0$ $\mathbf{to}$ $r_0-1$}{
  $\qcPsi_i\leftarrow$\textsc{circGen}$(n_0,m_s)$\;
  \For{$j\leftarrow 0$ $\mathbf{to}$ $r_0-1$}{
	\If{$\Pi_{i,j}=1$}{
	  $\qcM_{i,j}\leftarrow \qcPsi_{i}$\;
	}
   }
} 
}
\end{algorithm}

For the sake of simplicity, let us define $A^*=A\otimes 1_{p\times 1}$ and 
$B^*=B\otimes 1_{p\times 1}$.
We exploit the following result to obtain a strategy for performing an efficient
inversion of $Q$.

\textbf{\textit{Woodbury identity}}: Given two $n\times n$ matrices $W$ and $F$, where $F=UCL$, we have
\begin{equation}
(W+F)^{-1}=W^{-1}-W^{-1}U(C^{-1}+LW^{-1}U)^{-1}LW^{-1}.
\end{equation}

In the case of $\qcQ$, we have $W=\qcM$ and $F=\qcR$, so $U=A^*$, $C=I_z$ 
and $L=B^{*T}$. Using the Woodbury identity, we obtain
\begin{equation}
\label{eq:Qinv}
\qcQ^{-1}=\qcM^{-1}+\qcM^{-1}A^*\left(I_z+B^{*T}\qcM^{-1}A^*\right)^{-1}B^{*T}\qcM^{-1}.
\end{equation}
In order to facilitate the computation of $\qcQ^{-1}$, let us first consider that $\qcM^{-1}=\qcM^T$. We have
\begin{align}
\qcM^{-1}A^* & = Diag\left(\qcPsi^T_0,\qcPsi^T_1,\cdots,\qcPsi^T_{r_0-1}\right)\cdot\left(\Pi^T\otimes I_p\right)\cdot (A\otimes 1_{p\times 1})  \nonumber \\
 & = Diag\left(\qcPsi^T_0,\qcPsi^T_1,\cdots,\qcPsi^T_{r_0-1}\right)\cdot\left(\Pi^T\cdot A\right) \otimes  \left(I_p\cdot 1_{p\times 1}\right)\nonumber \\
 & = Diag\left(\qcPsi^T_0,\qcPsi^T_1,\cdots,\qcPsi^T_{r_0-1}\right) \cdot \left[\left(\Pi^T\cdot A\right)\otimes1_{p\times 1}\right] \nonumber \\
 & = \left(\Pi^T\cdot A\right)\otimes 1_{p\times 1}. 
\end{align}

The last equality is justified by the fact that the matrix $\left[\left(\Pi^T\cdot A\right)\otimes1_{p\times 1}\right]$ can be thought as the composition of 
vectors that having either the form of $0_{p\times 1}$ or $1_{p\times 1}$, 
and are thus invariant to permutations.
Hence, we have
\begin{align}
B^{*T}\cdot \qcM^{-1}\cdot A^* & = B^{*T}\cdot\left[\left(\Pi^T\cdot A\right)\otimes1_{p\times 1}\right]&\\\nonumber
& =\left(B^T\otimes1_{1\times p}\right) \cdot \left[\left(\Pi^T\cdot A\right)\otimes1_{p\times 1}\right]  \nonumber\\
& = \left( B^T\cdot\Pi^T\cdot A\right)\otimes \left(1_{1\times p}\cdot1_{p\times 1}\right).
\end{align}
The product $1_{1\times p}\cdot1_{p\times 1}$ corresponds to the sum of $p$ ones.
Therefore  it is equal to $1$ when $p$ is odd and equal to $0$ otherwise.
Based on these considerations, we obtain 
\begin{equation}
B^{*T}\qcM^{-1} A^*=\begin{cases}0_{z\times z} & \text{if $p$ is even}\\
B^T \Pi^T A&\text{if $p$ is odd},
\end{cases}
\end{equation}
and so we can define the matrix $D=I_z + B^{*T}\widetilde{M}^{-1}A^*$, such that
\begin{equation}
D=\begin{cases}I_z &\text{if $p$ is even,}\\
I_z + B^T \Pi^T A &\text{if $p$ is odd.}\end{cases}
\end{equation}
So, combining these results, regardless of the parity of $p$, we have that
\begin{equation}
\qcQ^{-1}=\qcM^T+\qcM^T A^* D^{-1}B^{*T}\qcM^T.
\end{equation}
This expression can be further simplified by considering the special structure 
of the involved matrices, thus obtaining the following expression for $\qcQ^{-1}$ 
that is convenient from the complexity standpoint:
\begin{equation}
\label{eq:Qinv_even}
\qcQ^{-1} = \qcM^T+\left(\Pi^T\cdot A \cdot B^T\cdot\Pi^T\right)\otimes 1_{p\times p}.
\end{equation}
We report the full derivation in Appendix~\ref{app:qcQ}.

Based on this analysis, we note that the choice of an even $p$ 
simplifies the computation of $\qcQ$ and $\qcQ^{-1}$, since it guarantees that 
$\qcQ$ can always be inverted because $D=I_z$.
However, it has been recently shown that using circulant blocks with even size 
may reduce the security of the systems relying on them~\cite{Londahl2016}.
Therefore, it is advisable to choose odd values of $p$, although in this case 
the non-singularity of $\qcQ$ is no longer guaranteed and more than one attempt 
may be needed to generate a non-singular $\qcQ$.
We point out that, in the case of $z$ assuming small values (such as the 
ones we consider in this paper) this choice has a negligible impact on the 
efficiency of the scheme, since generating $D$ and checking its non-singularity 
is fast.

\subsection{Public key generation}

The public key is simply computed as
\begin{equation}
\qcH'=\qcQ^{-1}\qcH\qcS^{-1}.
\end{equation}
Exploiting the systematic structure of $\qcH$, we have
\begin{align}
\label{eq:public_key}
\qcH'=\begin{bmatrix}\left.\qcQ^{-1} \qcV^T\right|\qcQ^{-1}\end{bmatrix}\qcS^{-1}.
\end{align}
\subsection{Signature Generation}

In order to implement the function $\mathcal{F}_\Theta$ introduced in Section 
\ref{sec:description}, let us consider a constant weight encoding function 
$CW(d,n,w)$ that takes as input a binary vector $d$ of given length and returns 
a length-$n$ vector with weight $w$. 
In particular, given a message $m$ that must be signed, we choose 
$d = \mathcal{H}(m)$, where $\mathcal{H}$ is a public hash function.
The input given to the constant weight encoding function is the concatenation of the 
digest $d$ with the binary representation of the parameter $\Theta$, which can be the 
value of a counter or a pseudo-random integer variable. 
In other words, given a message $m$, the parameter $\Theta$ is used to obtain several 
different outputs from the constant weight encoding function.
This feature is necessary because, as we explain in section \ref{sec:err_vec}, the output 
of $\mathcal{F}_\Theta$ must be in the kernel of $\qcR$.
If, for a given $\Theta$, the current output does not verify this property, we just 
change the value of $\Theta$ and try again.
The signature of a message $m$ is constituted by the binary string 
$\sigma=(e + c) \qcS^T$ and the chosen value of $\Theta$, denoted as $\Theta^*$. 
In the signature generation, $c$ is a codeword of the code $\mathcal{C}$ with 
weight $w_c$, and $e$ is an error vector with weight $w$ which is generated 
from $m$ and $\Theta^*$ through the function $CW$.

\subsubsection{Random codeword generation}

Each signature is built upon a random sparse codeword $c$, with weight $w_c\ll n$. 
As we explained in Section \ref{subsec:codingbackground}, such a codeword can
be easily obtained by choosing $w_c = m_g w_g$, with $m_g$ being a small integer.
Let $u$ be the length-$k$ information sequence corresponding to $c$; 
thanks to the systematic form of $\qcV$, we have
\begin{equation}
\label{eq:codeword}
c=u\qcG = \left[ \left. u \right| u\qcV \right].
\end{equation}
This means that we can easily obtain such a codeword randomly picking a 
sequence $u$ of weight $m_g$ and computing the corresponding codeword as in eq. \eqref{eq:codeword} picking a different set of codewords to be added together
if the weight of the sum does not fit.

\subsubsection{Error vector generation\label{sec:err_vec}}
In order to generate the error vector $e$, we first compute its syndrome 
$s$ as the digest of the message $m$, through the function $\mathcal{F}_\Theta$, 
by choosing a value of $\Theta = \Theta^*$ such that the product $s' = \qcQ s$ 
has the same weight of $s$.
Subseuently, the error vector is obtained as $e=\left[ \left. 0_{1\times k_0p} 
\right| s'^T \right]$.
We point out that the constraint on the weight of $s'$ can be simply satisfied by imposing $\left( B^T \otimes 1_{1 \times p} \right) s = 0$. 
Indeed, recalling Eq. \eqref{eq:RM}, we have
\begin{align}
s' \nonumber & = \qcQ s = \qcM s + \qcR s  \\\nonumber
& = \qcM s + A^* B^{*T} s  \\\nonumber
& = \qcQ s + (A\otimes 1_{p\times 1})(B^T\otimes 1_{1\times p}) s.
\end{align}
Since $\qcM$ is a permutation matrix, when the product $\widetilde{R} s$ is null, 
$s'$ just corresponds to a permuted version of $s$.
This condition can be checked efficiently. First of all, 
let us write $s=[s^{(p)}_0, s^{(p)}_1, \cdots, s^{(p)}_{r_0-1}]^T$, where each 
$s^{(p)}_i$ is a length-$p$ sequence.
In the same way, we write $s'=[s'^{(p)}_0, s'^{(p)}_1, \cdots, s'^{(p)}_{z-1}]^T$.
Through some straightforward computations, it can be verified that $s_i'^{(p)} = 
0_{p\times 1}$ only when the sum of the Hamming weights of the blocks $s_j^{(p)}$ 
indexed by the $i$-th row of $B^T$ is even. 

The syndrome $s$ is constructed from $m$ through $CW$ and has fixed weight equal 
to $w$.
An algorithmic way to compute the syndrome $s$ and the corresponding error 
vector $e$ is described in Algorithm \ref{alg:genE}.
A parameter to optimize is $\Theta_{\max}$, representing the maximum value allowed for $\Theta$, which must be sufficiently large to ensure that a value $\Theta = \Theta^*$ such that $B^{*T} s = 0$ is found with very high probability.
Thus, by increasing $\Theta_{\max}$, the probability of a signing failure can 
be made negligible.

To this end, in the implementation we chose to represent the counter $\Theta$ as 
a $64$-bit unsigned integer value.
This limits the probability $\left(1-\frac{1}{z}\right)^{\Theta_{\max}}$ of 
never finding a $64$-bit value $\Theta^*$ such that $B^{*T} s = 0$ to 
$2^{-512}$ for $z$ up to $50$. We remark that the current parametrization for 
the proposed \sysacro{} primitive employs $z = 2$, thus making the failure 
probability negligible for all practical purposes.
Once the error vector is obtained, the signature is computed as $\sigma = (e + c) \widetilde{S}^T$.

\begin{algorithm}[!t]
{\small 
\LinesNumbered
\DontPrintSemicolon
\caption{Generation of $e$\label{alg:genE}}
\KwIn{$b$, $M^T$, $m$, $w$, $\Theta_{\max}$}
\KwOut{$\Theta^*,e$}
$w_s=1$\;
\While{$\left\{w_s>0\right\}$} {
    $\Theta\leftarrow$\textsc{randGen}$(\Theta_{\max})$\;
	$d\leftarrow \mathcal{H}\left([m|\Theta]\right)$\;
	$s\leftarrow CW(d,w,r_0p)$\;
	$i\leftarrow 0$\;
	\While{$ i < z $} {
		$w_s \leftarrow 0$\;
		\For{$j \leftarrow 0$ $\mathbf{to}$ $r_0-1$} {
			\If{$B^T_{j,i}=1$}{
			   $w_s\leftarrow w_s+wt(s_{j})$\;
			}
	    }
		$w_s\leftarrow w_s \bmod 2$\;
		\If{$w_s=1$}{
			$i\leftarrow z$\;
        }    
		\Else {
		   $i\leftarrow i+1$
		}
	}
}
$s'\leftarrow \widetilde{M}s$\;
$e\leftarrow \left[0_{1\times k_0p} | s'^T\right]$\;
$\Theta^* \leftarrow \Theta$\;
}
\end{algorithm}

\subsubsection{Number of different signatures}

An important parameter for any digital signature scheme is the total number of 
different signatures. 
Computing such a number is useful, for example, to verify that collision attacks 
are unfeasible (see Section \ref{sec:OtherAttacks}).
In \sysacro{}, a unique signature corresponds to a specific $r$-bit vector 
$s$, having weight $w$.
Only vectors $s$ being in the kernel of $\widetilde{R}$ are acceptable: since $\widetilde{R}$ has rank equal to $z$, then its kernel has dimension $r-z$, 
which means that the number of binary vectors being in its kernel is equal to 
$2^{r-z}$.
We suppose that these $2^{r-z}$ vectors are uniformly distributed among 
all the vectors of length $r$.
This in turn implies that, considering the $r$-tuples with weight $w$, we 
expect a fraction $2^{-z}$ of them to be in the kernel of $\widetilde{R}$. 
Thus, the total number of different signatures is
\begin{equation}
N_s \approx \frac{\binom{r}{w}}{2^z}.
\label{eq:signum}
\end{equation}

\subsection{Signature verification}

According to \cite{Baldi2013c}, the signature generation basically coincides 
with the computation of a new syndrome through the public code and the execution 
of some checks on  the result.
The final check consists in verifying that the new syndrome coincides with the 
one resulting from feeding the message digest to the $CW$ function.
These two vectors should coincide because
\begin{align}
\widetilde{H}'\sigma^T&=\widetilde{H}'\widetilde{S}(c^T+e^T)\nonumber \\
&=\widetilde{Q}^{-1}\widetilde{H}\widetilde{S}^{-1}\widetilde{S}(c^T+e^T)\nonumber \\
&=\widetilde{Q}^{-1}\widetilde{H}(c^T+e^T)\nonumber \\
&=\widetilde{Q}^{-1}\widetilde{H}e^T\nonumber \\
&=\widetilde{Q}^{-1}\widetilde{Q}s\nonumber \\
&=s.
\end{align}

An algorithmic description of the signature verification procedure is reported in Algorithm \ref{alg:sigVer}.

\begin{algorithm}[!t]
{\small 
\LinesNumbered
\DontPrintSemicolon
\caption{Verification of $\sigma$\label{alg:sigVer}}
\KwIn{$H'$, $\sigma$, $m$, $\Theta^*$}
\KwOut{$ans$: Boolean value indicating if the signature is verified or not}
\BlankLine	
$w_{\sigma}\leftarrow wt(\sigma)$ \;
\If {$w_{\sigma}>(w+m_g w_g)m_s$}{
    \KwRet{\texttt{false}}
}
$d^*\leftarrow \mathcal{H}\left([m|\Theta^*]\right)$\;
$s^*_1\leftarrow CW\left(d^*,r_0p,w\right)$\;
$s^*_2\leftarrow \qcH'\sigma^T$\;
\If{$s^*_1\neq s^*_2$}{
   \KwRet{\texttt{false}}
}   
\Else{
   \KwRet{\texttt{true}}
}
}
\end{algorithm}
%
%
\section{Security Analysis}\label{sec:security}
In this section we review the main known attack strategies against \sysacro{} 
and their complexity.

\subsection{Decoding attacks}
\label{sec:ISD}

In \sysacro{}, an attacker knows that $\sigma = (e+c) \cdot \qcS^T = e'' + c''$, 
with $c''=c\qcS^T$ being a codeword of the public code, i.e, such that $\qcH' \cdot c''^T = 0$.
Hence, $e'' = e\qcS^T$ can be considered as an error vector with weight $\le m_s 
w$ affecting the codeword $c''$ of the public code and having $s$ as its 
syndrome.
Therefore, an attack strategy consists in exploiting decoding algorithms for general linear codes to recover $e''$ from $s$ and $\qcH'$. If this succeeds, then the attacker has to find a codeword $c''$ of the public code with suitable weight to be added to $e''$ in order to forge a valid signature.

The problem of finding $e''$ from $s$ is known as \ac{SDP}.
If the \ac{SDP} admits a unique solution, \ac{ISD} algorithms are those 
achieving the best performance in solving it.

In order to determine whether the \ac{SDP} has a unique solution or not, we need to estimate the minimum distance of the public code.
The public code admits a generator matrix in the form $\qcG'=\qcG\cdot \qcS^T$, which is also sparse.
Hence the public code contains low weight codewords, coinciding with the rows of $\qcG'$, which have weight approximately equal to $w_g m_s$.
Since the sum of any two or more rows gives a codeword with weight $\geq w_g m_s$ with overwhelming probability, we can take $w_g m_s$ as a reliable estimate of the minimum distance of the public code, which will hence be characterized by decoding spheres with radius  $t^*=\left\lfloor \frac{w_g m_s-1}{2}\right\rfloor$.
Since we want to guarantee the uniqueness of the \ac{SDP} solution, we must impose
\begin{equation}
 m_s w \leq t^*.
\end{equation}
In order to satisfy the previous condition, we choose $w_g=2 w + 1$, leading to $t^* = w m_s + \left\lfloor\frac{m_s-1}{2}\right\rfloor$.
With this choice, we guarantee that there is no algorithm that can solve the \ac{SDP} more efficiently than \ac{ISD}, thus we consider the \ac{ISD} \ac{WF} to compute the security level of \sysacro{} against decoding attacks.

The \ac{ISD} approach, which was pioneered by Prange in~\cite{Prange1962}, 
attempts at performing the decoding of a general linear code more efficiently 
than an exhaustive search approach.
Subsequent improvements of Prange's algorithm were presented by Lee and 
Brickell~\cite{Lee1988}, Leon~\cite{Leon1988} and Stern~\cite{Stern1989}.
Among these variants, Stern's algorithm~\cite{Stern1989} is currently the one best exploiting the speedups provided by quantum computers, as shown in~\cite{Vries2016}.
In particular, a significant portion of Stern's algorithm can be solved
employing Grover's algorithm~\cite{Grover1996} to reduce the running time
to the square root of the one needed for the computation on a classical platform.
By contrast, when execution on classical computers is considered, the most 
efficient \ac{ISD} turns out to be the Becker-Joux-May-Meurer (BJMM) algorithm 
proposed in~\cite{Becker2012}, which is part of a family of results on the 
subject~\cite{May2011,Peters2010,Bernstein2011,Niebuhr2017}.
All the aforementioned approaches have a running time growing exponentially
in the effective key size of the scheme (a function of the number of
errors, code size and rate), regardless of the availability of a quantum computer.

As a consequence, the security levels against attackers performing a \ac{DA} with 
classical computers have been estimated by considering the \ac{WF} of the BJMM 
algorithm, while the security levels against quantum computer-equipped attackers 
were computed taking into account Stern's algorithm.

We defend \sysacro{} from \acp{DA} employing parameters which prevent the 
syndrome decoding from succeeding given a computational power bounded by the desired security level.
To this end, we take into account the fact that the nature of the \ac{QC} 
codes employed in \sysacro{} provides a speedup by a factor $\sqrt{p}$ with respect to the running time of the \ac{ISD} algorithm employed to perform decoding of a general linear code~\cite{Sendrier2011}.

\subsubsection{Quantum Stern's algorithm}
\label{subsec:PQStern}

Considering the fact that Stern's algorithm~\cite{Stern1989}
is the one best suited for quantum computer execution, and will thus
be employed to determine the parameters of \sysacro{}, 
we briefly resume the results in~\cite{Vries2016}, describing 
how the application of Grover's algorithm to \ac{ISD} algorithms can be taken into account when computing the complexity of \acp{KRA} and \acp{DA}.

An \ac{ISD} algorithm is an algorithm $\mathcal{A}\left( \mathcal{C}(n,k),w\right)$ 
taking as input a code $\mathcal{C}(n,k)$ with length $n$, dimension $k$, and 
trying to find a codeword with weight $w$ or, equivalently, an error vector with 
weight $w$ given the code and the corresponding syndrome of the error through the 
code.
In \sysacro{}, employing \ac{ISD} to perform general decoding we have it acting 
on an $n_0p$ bits long code, with dimension $k_0p$, trying to correct an error 
vector with weight $\leq m_s w$.

The basic structure of each \ac{ISD} algorithm is essentially the same, and 
relies on the identification of an \textit{information set}, that is, guessing a set of error-free positions in the error-vector, corresponding to a set of 
$k$ linearly independent columns of the generator matrix of the code.
Recovering the entries of the error vector affecting this set is enough to reconstruct the whole error vector.
The algorithm must be run iteratively, and each iteration has a probability of success $p_{\mathcal{A}}$.
Thus, the expected number of iterations that makes the attack successful is 
$\frac{1}{p_{\mathcal{A}}}$.
The probability $p_{\mathcal{A}}$ is obtained as the product of $p_{inv}$ and $p_e$, where $p_{inv}$ is the probability that an iteration of \ac{ISD} has selected a set of $k$ linearly independent vectors, while $p_{e}$ is the probability that the error vector entries affecting the selected set can be recovered.
It can be proven that $p_{inv}$ converges to $p_{inv}\approx 0.29$ as the size of the binary matrix being inverted increases \cite{Vries2016}, while for $p_{e}$ we have
\[
p_{e}=\frac{\binom{w}{2j}\binom{n-w}{k-2j}\binom{2j}{j}\binom{n-k-w+2j}{l}}{4^j\binom{n}{k}\binom{n-k}{l}}
\]
according to~\cite{Stern1989}, where $l$ and $j$ are parameters which influence the complexity of the algorithm and must be optimized to minimize the value of $p_e$.

Taking into account the speedup following from the application of Grover's algorithm to Stern's algorithm, it follows that the algorithm is successful after performing only 
$\frac{\pi}{4} \sqrt{\frac{1}{p_{\mathcal{A}}}}=\frac{\pi}{4}\sqrt{\frac{1}{p_{inv}p_e}}$ iterations on average, instead of 
$\frac{1}{p_{inv}p_e}$.
Let us define:
\begin{itemize}
\item $c_{dec}$ as the cost in qubit operations of decoding the input qubits to the inputs of the classical algorithm which must be performed whenever an iteration is completed on the quantum computer;
\item $c_{it}$ as the number of bit operations needed to perform an iteration of the classical Stern's algorithm;
\item $c_{inv}$ as the cost of inverting the matrix obtained with the $k$ columns selected during the iteration; in fact, since a quantum implementation of Stern's algorithm must be performed entirely with revertible operations, skipping an iteration is not possible, even if the selected $k$ columns do not correspond to an information set (i.e., they are not linearly independent).
\end{itemize}
By taking the conservative assumption that a qubit operation has the same cost of a bit operation, it is possible to express the amount of operations required to execute Stern's algorithm on a quantum computer as
\begin{equation}
\label{eq:PQStern}
\frac{\pi}{4}\sqrt{\frac{1}{p_{inv}p_e}}(c_{dec}+c_{inv}+c_{it}).
\end{equation}

Estimating the actual value of $c_{dec}$ may be very hard, since it depends on the size of the input given to $\mathcal{A}$.
For example, some input parameters can be fixed (in this case, the number of bits needed to represent the input given to $\mathcal{A}$ decreases) but, at the same time, the value of $p_e$ might get lower (since, in this case, we might not consider an optimal input choice).
While estimates for $c_{dec}$ have put it in the $2^{30}$ range~\cite{Vries2016}, we conservatively consider $c_{dec}=0$.
Finally, to compute the two remaining computational costs, we refer to the following expressions (from~\cite{Stern1989})
\begin{equation}
c_{it}=2lj\binom{k/2}{j}+2j(n-k)\binom{k/2}{j}^2 2^{-l},
\end{equation}
\begin{equation}
c_{inv}=\frac{1}{2}(n-k)^3+k(n-k)^2.
\end{equation}
We point out that, for the cases we are interested in, the values of \eqref{eq:PQStern} slightly depend on $c_{dec}$, so we can conservatively neglect it, without significant variations in the attack complexity.

\subsubsection{BJMM algorithm complexity.}
\label{subsec:BJMM}

As already mentioned, when only classical computers are available, the most 
efficient \ac{ISD} algorithm turns out to be the BJMM algorithm proposed in 
~\cite{Becker2012}.
A precise estimate of the \ac{WF} of this algorithm in the finite-length 
regime can be found in \cite{Hamdaoui2013}, and it has been used to compute the 
\ac{WF} of attacks based on \ac{ISD} against the proposed instances of 
\sysacro{}, when performed with classical computers.
While the complete expression of the computational complexity of the BJMM algorithm is
rather involved, we point out that a simple expression providing an approximate 
but fairly
intuitive expression for it is reported in~\cite{CantoTorres2016} and is $2^{cw}$, where $c=\log_2{\frac{1}{1-\frac{k}{n}}}$.

\subsection{Key recovery attacks}
\label{sec:KeyRecAtt}

An attacker could aim to mount a \ac{KRA} against \sysacro{}, aimed at obtaining the private key.
A potential vulnerability in this sense comes from the use of LDGM codes: these codes offer the advantage of having a predictable (and sufficiently
high) number of codewords with a moderately low weight $w_c$, and of making their random selection very easy for the signer.
On the other hand, as pointed out in the previous section, the public code is characterized by low weight codewords with weight approximately equal to $w_g m_S$.
Since $\qcG'$ has $k$ rows, and summing any two of them gives higher weight codewords
with overwhelming probability, we can consider that the multiplicity of these
low weight codewords in the public code is $k$.

It is possible to show that the low weight codeword finding problem is equivalent to the general linear code decoding problem, thus allowing \ac{ISD} to be retrofit to this task too. 
Thus, rows of $\qcG'$ might be recovered using \ac{ISD} algorithms to search for codewords of weight $\leq w_gm_S$ in the public code.

We assume that knowing one row of $\qcG$ can be enough to recover the whole matrix, even if this is a very conservative approach. 
Taking into account the multiplicity of low weight codewords, we consider a speedup factor of $k$, with respect to the application of \ac{ISD} to a general code.

As already explained in Section \ref{sec:ISD}, in the case of a quantum computer-equipped attacker, the best ISD algorithm is Stern's algorithm, (described in Section \ref{subsec:PQStern}), while in the case of a classical computer, the best solution is the BJMM algorithm (described in Section \ref{subsec:BJMM}).

Another possible vulnerability comes from the fact 
that an attacker could obtain the vector space generated by $B^T\otimes 1_{1\times p}$, as well as its dual space,
by observing $O(r)$ public syndromes $s$, since $B^T\otimes 1_{1\times p} \cdot s = 0_{z \times 1}$.
Hence, we can suppose that an attacker knows an $r \times r$ matrix $U$ such that $\qcR \cdot U = 0 \Rightarrow \qcQ \cdot U = \qcM \cdot U$.
The attacker also knows that $\qcH' = \qcQ^{-1} \cdot \qcH \cdot \qcS^{-1}$ and that the public code admits any non-singular generator matrix in the form
$G_X' = X \cdot \qcG \cdot \qcS^T$, which becomes $G_{\qcQ}' = \qcQ \cdot \qcG \cdot \qcS^T$ for $X = \qcQ$.
The matrix $G_I'$, which corresponds to the choice of $X=I$, is likely to be the most sparse one among them, and it can be attacked by searching for low weight codewords in the public code,
as we have already observed.
On the other hand, knowing $V$ does not help to reduce the complexity of attacking either $\qcH'$ or one of the possible $G_X'$,
hence it cannot be exploited by an attacker to perform a \ac{KRA}.
A newly devised attack which is more efficient in recovering the secret key is instead targeted against
the matrix $\qcS^T$, and will be discussed in Section~\ref{sec:StatAttacks}.

\subsection{Collision attacks}
\label{sec:OtherAttacks}

As for any other hash-and-sign scheme, classical collision birthday attacks represent a threat for \sysacro{}.
Since the system admits up to $N_s$ different signatures, where $N_s$ is given by \eqref{eq:signum}, it is sufficient to collect $\approx \sqrt{N_s}$ different signatures to have a high probability of finding a collision with a classical computer.
Hence, the security level reached by the system under classical computing cannot exceed $\sqrt{N_s}$.

If we consider an attacker provided with a quantum computer, we must take into account the BHT algorithm \cite{Brassard1998}, implying that the security level cannot exceed $\sqrt[3]{N_s}$.

\subsection{Forgery attacks based on right-inverse matrices}
\label{sec:ForgeryAttacks}

In order to forge signatures, an attacker could search for an $n \times r$ right-inverse matrix $\qcH'_r$ of $\qcH'$. 
If the signature were dense, it would have been easy to find a right-inverse matrix able to forge it.
In fact, provided that $\qcH' \cdot \qcH'^T$ is invertible, then
$\qcH'_r = \qcH'^T \cdot \left(\qcH' \cdot \qcH'^T\right)^{-1}$ is a right-inverse matrix of $\qcH'$.
The matrix $\qcH'$ is dense and the same occurs, in general, for $\left(\qcH' \cdot \qcH'^T\right)^{-1}$.
So, $\qcH'_r$ is dense as well.
Then, the attacker could compute $f = (\qcH'_r \cdot s)^T$, which is a forged signature, and will be dense with overwhelming probability.

However, the right-inverse matrix is not unique.
So, the attacker could search for an alternative, possibly sparse, right-inverse matrix.
In fact, given an $n \times n$ matrix $Z$ such that $\qcH' \cdot Z \cdot \qcH'^T$ is invertible,
$H''_r = Z \cdot \qcH'^T \cdot \left(\qcH' \cdot Z \cdot \qcH'^T\right)^{-1}$
is another valid right-inverse matrix of $\qcH'$.
We notice that $H''_r \ne Z \cdot \qcH'_r$.
When $\qcH'$ contains an invertible $r\times r$ square block, there is also another simple way to find 
a right-inverse.
It is obtained by inverting such 
block, putting its inverse at the same position (in a transposed matrix)
in which it is found within $\qcH'$, and padding the remaining rows with zeros.

In any case, there is no simple way to find a right-inverse matrix that is also sparse, 
which is the aim of an attacker.
Actually, for the matrix sizes considered here, the number of possible 
choices of $Z$ is always huge.
Moreover, there is no guarantee that any of them produces a sparse right-inverse.
Searching for an $r \times r$ invertible block within $\qcH'$ and inverting it would also
produce unsatisfactory results, since the overall density of $\qcH'^{-1}$ is reduced,
but the inverse of the square block is still dense.

So, the attacker would be able to forge signatures with a number of symbols $1$ in the order of $r/2$, that is still too large for the system considered here.
In fact, in all the proposed \sysacro{} instances, the density of signatures is in the order of $1/3$ or less.

One could also think of exploiting \ac{ISD} or low-weight codeword searching algorithms to find a sparse representation of the column space of $\qcH'_r$. If this succeeds, it would result in a sparse matrix $H_W = \qcH'_r \cdot W$, for some $r \times r$ transformation 
matrix $W$.
However, in this case, $H_W$ would not be a right-inverse of $\qcH'$.

For the above reasons, approaches based on right-inverses are not feasible for an attacker.

\subsection{Forgery attacks based on linear combinations of signatures}
\label{LinearCombinationAttack}

As another attack strategy, an opponent might sum a number $L$ of syndromes $\left\{s^{(1)},s^{(2)},\cdots,s^{(L)}\right\}$ and obtain an artificial syndrome $s^*$ with weight $w$.
By the sum of the corresponding signatures, he could obtain a signature $\sigma^*$ such that $s^*=\qcH' \cdot \sigma^{*T}$. 
However, $\sigma^*$ can be accepted only if it has weight $\leq (w + w_c) m_s$.
Since $\sigma^*$ is obtained as the sum of sparse vectors, with very high probability its density is higher than that of the added signatures.
An attacker must hence look for special linear combinations of valid signatures which produce forged signatures having an acceptably low weight.
We denote this kind of attack as linear combination attack (LCA) and estimate its \ac{WF} as follows.

First of all, let us consider the following fact: the signature can be written as $\sigma=[c+e] \cdot \qcS^T=\left[u|u \cdot \qcV+s \cdot \qcQ^T\right] \cdot \qcS^T$, where $u$ is a random information sequence having weight $m_g$.
After the opponent finds a set of $L$ syndromes leading to an $s^*$ with the desired weight, he computes the corresponding $\sigma^*$ as
\begin{align}
\sigma^* &= \left[\left.\sum_{i=1}^{L}{u^{(i)}}\right|\sum_{i=1}^{L}{u^{(i)}}\qcV+s^* \cdot \qcQ^T\right] \cdot \qcS^T \nonumber \\
&= \left[\left.u^*\right|u^* \cdot \qcV+s^* \cdot \qcQ^T\right] \cdot \qcS^T.
\end{align} 
If $u^*$ has weight $\leq m_g$, then the forged signature will have the desired weight.
Since the choice of vector $u$ is independent of the syndrome $s$, we have
\begin{equation}
P\left\{wt\left(s^*\right)=w,\hspace{0.5mm} wt\left(e^*\right)\leq w_{\sigma} \right\} = P\left\{wt \left(s^*\right)=w \right\}\cdot P\left\{ wt\left(u^*\right)\leq m_g \right\}
\end{equation}
where 
$w_{\sigma}=(w+w_c)m_s$.
If we define $P_{\oplus}(N,n,w_x,w_y)$ as the probability that the XOR of $N$ different length-$n$ and weight-$w_x$ vectors results in a vector with weight $w_y$, we have
\begin{equation}
P\left\{wt \left(s^*\right)=w \right\}=P_{\oplus}\left(L,r,w,w\right), 
\end{equation}
\begin{equation}
P\left\{ wt\left(u^*\right)\leq m_g \right\}=\sum_{x=0}^{m_g}{P_{\oplus}\left(L,k,m_g,x\right)}.
\end{equation}
The expression of $P_{\oplus}(N,n,w_x,w_y)$ can be found in Appendix \ref{app:PSIA}.

The \ac{WF} of this attack can be computed as
\begin{equation}
\label{eq:WF_forge}
WF_{\mathrm{LCA}}=\frac{C_s(L)+C_{\sigma}(L)}{P\left\{wt\left(s^*\right)=w,\hspace{0.5mm} wt\left(e^*\right)\leq w_{\sigma} \right\}}
\end{equation}
where $C_s(L)=(L-1)w$ is the number of operations needed to sum the $L$ syndromes and $C_{\sigma}(L)=(L-1)w_{\sigma}$ is the cost of summing the signatures.
The opponent is free to choose the value of $L$ which minimizes the attack complexity.
For all the proposed \sysacro{} instances, the attack is optimized by choosing $L=2$.

Concerning the possible applicability of quantum speedups to forgery attacks based on linear combinations of syndromes, there is no apparent way of applying Grover's algorithm to accelerate these attack procedures.
Therefore, it is reasonable to assume that no quantum speedups apply, i.e., $WF^{(pq)}_{\mathrm{LCA}} = WF^{(cl)}_{\mathrm{LCA}} = WF_{\mathrm{LCA}}$ as expressed by eq. \eqref{eq:WF_forge}.
However, in order to provide a conservative option, in the proposed instances of \sysacro{} we also consider the theoretical case in which the maximum possible speedup due to Grover's algorithm is achieved, i.e., $WF^{(pq)}_{\mathrm{LCA}} = \sqrt{WF^{(cl)}_{\mathrm{LCA}}} = \sqrt{WF_{\mathrm{LCA}}}$.

\subsection{Forgery attacks based on support intersections}
\label{sec:SIA}

If we suppose that $e$ and $c$ have disjoint supports, the effect of the
proposed scheme on the public syndrome $s$
can be seen as the expansion of an $r \times 1$ vector with weight $w$ into
a subset of the support of the $1 \times n$ vector $\sigma$, having weight $\le m_s w$,
in which each symbol $1$ (set symbol) in $s$ corresponds, at most, to $m_s$ symbols $1$ in $\sigma$.

An attacker could try to find the $w$ sets of $m_s$ (or less) symbols $1$ within 
the support of $\sigma$ to be able to compute valid signatures.
In this case, he will work as if the random codeword was absent, that is, $c = 0_{1 \times n}$.
Thus, even after succeeding, he would be able to forge signatures that are sparser
than the authentic ones.
In any case, this is a potential weakness, so we design the system in such a way as to avoid its occurrence.

To reach the target, the attacker must collect a sufficiently large number $L$ of pairs $(s, \sigma)$. 
Then, he can intersect the supports (that is, compute the bit-wise AND) of all the vectors $s$.
This way, he obtains a vector $s_L$ that may have a small weight $w_L \ge 1$.
If this succeeds, the attacker analyzes the vectors $\sigma$, and selects the $w_L' = m_s w_L$
set bit positions that appear more frequently.
If these bit positions are actually those corresponding to the $w_L$ bits set in $s_L$,
then the attacker has discovered the relationship between them. In principle, it is convenient for the attacker to choose the smallest possible value of $w_L$. 
However, for this analysis to be successful it is necessary that $w_L \geq d$, where $d$ is the minimum distance of the code defined by the matrix $B$ introduced in \eqref{eq:RM} \cite{Baldi2013c}.
A general expression of the probability of success of a SIA is derived next.
Let us denote as $I$ the set of $w'_L$ bits in $e'$ that correspond to the $w_L$ bits set in $s_L$, and as $J$ the complement of $I$.
The probability that an attacker finds a vector $s_L$ with weight
$w_L$ and discovers the $w'_L$ bits corresponding to the $w_L$ bits set
in $s_L$ can be computed as
\begin{equation}
P_{\mathrm{SIA}} = P_\land(r, \left\{s_1, s_2, \ldots, s_L\right\}, w_L) \cdot P_{I \ge J,L}(w_L, w_c,w'_L),
\label{eq:PSIA}
\end{equation}
with the functions $P_\land(\bullet)$ and $P_{I \ge J,L}(\bullet)$ having the expressions reported in Appendix \ref{app:PSIA}, where the proof of \eqref{eq:PSIA} is also reported. 
Equation \eqref{eq:PSIA}, however, is not yet sufficient to compute the \ac{WF} of a SIA. This is because the attacker should repeat the procedures so many times as to have a group of relationships able to cover all possible syndrome vectors with weight $w$. The implications of this fact are discussed next.

Let us suppose that $w_L$ divides $w$. After having computed the public syndrome of the message, the attacker intersects vectors $s$ until he obtains a vector $s_L$ with weight $w_L$ having symbols $1$ in correspondence of $w_L$ out of the $w$ positions. The probability that this happens is
\begin{equation}
P_\land(r, \left\{s_1, s_2, \ldots, s_L\right\}, w_L) \frac{\binom{w}{w_L}}{\binom{r}{w_L}}. \nonumber
\label{eq:PSIA_a}
\end{equation}

In order to recover the positions of all the $w$ symbols $1$ of the syndrome, this procedure must be repeated $\frac{w}{w_L}$ times, at any step $i$ ($i = 1, . . ., \frac{w}{w_L}$) involving $w - (i - 1)w_L$ positions. Taking into account that the attack at any step is successful if the bits at positions $\in I$ are set more times than the bits at positions $\in J$, the probability of success at the $i$-th step results in
\begin{equation}
P_{\mathrm{SIA}}^{(i)} = P_\land(r, \left\{s_1, s_2, \ldots, s_L\right\}, w_L) \cdot \frac{\binom{w - (i - 1)w_L}{w_L}}{\binom{r}{w_L}} P_{I \ge J,L}(w_L, w_c,w'_L).
\label{eq:PSIA_b}
\end{equation}

When $w_L$ does not divide $w$, the attacker can follow two strategies.
According to the first strategy, the procedure is repeated $\lfloor \frac{w}{w_L} \rfloor$ times by assuming a greater value of $w_L$ in the last step.
According to the second strategy, the procedure is repeated $\lceil \frac{w}{w_L} \rceil$ times and, in the last step, the attacker intersects vectors $s$ until he obtains a vector $s_L$ with weight $w_L$ having $w - \left( \lceil \frac{w}{w_L} \rceil - 1 \right)w_L < w_L$ set symbols in correspondence of as many positions, and the remaining $w_L - w + \left( \lceil \frac{w}{w_L} \rceil - 1 \right)w_L$ set symbols in correspondence of previously selected positions (i.e., during previous steps, the attacker must identify these positions). It is easy to verify that the probability that such a vector $s_L$ is found results in
\begin{equation}
P_\land(r, \left\{s_1, s_2, \ldots, s_L\right\}, w_L) \frac{\binom{\lceil w/w_L \rceil - 1)w_L }{w_L - w + \lceil w/w_L \rceil - 1)w_L}}{\binom{r}{w_L}}.
\label{eq:PSIA_c}
\end{equation}

\subsubsection{Work factor of a SIA}
\label{subsec:SIAWF}

A realistic estimate of the $WF$ of a SIA can be found as follows.
When $w_L$ divides $w$, the $WF$ for the $i$-th step of the SIA can be written as
\begin{equation}
WF_{\mathrm{SIA}}^{(i)} = \left\{ \left[ \frac{C_{s_L,1}}{P_\land(r, \left\{s_1, s_2, \ldots, s_L\right\}, w_L)} + C_{s_L,2} \right] \cdot \frac{\binom{r}{w_L}}{\binom{w - (i - 1)w_L}{w_L}} + C_{I \ge J,L} \right\} \frac{1}{P_{I \ge J,L}(w_L, w_c,w'_L)}. 
\label{eq:WFSIA}
\end{equation}
In this expression, $C_{s_L,1} = (L - 1)w$ denotes the number of binary operations needed to intersect $L$ syndrome vectors, $C_{s_L,2} = w_L$ denotes the number of operations required to check if the $w_L$ symbols $1$ of the vector $s_L$ are in the desired positions, and $C_{I \ge J,L} = (L-1) n \lceil \log_2(L) \rceil + w'_Lr$ denotes the number of operations needed to find the $w_L'$ bits of $e$ that should correspond to the $w_L$ symbols $1$ of $s_L$ and verify whether the attack has been successful or not.

More precisely, the first term counts the number of bits set in each position of a group of $L$ collected signatures, i.e., $(L - 1)n$ sums, each requiring $\lceil \log_2(L) \rceil$ bits to store the result (since the sum is not binary).
The second term counts the operations needed to perform the verification step
(that is, a vector-matrix multiplication requiring $w'_Lr$ binary operations).
In fact, once having found the $w'_L$ set bit positions that appear more frequently within 
the $L$ signatures $\sigma$, the attacker must check if they actually correspond
to the $w_L$ bits set in $s_L$. For this purpose, he can simply check if $s_L = H' \cdot \sigma_L$, where $\sigma_L$
is the $n$-bit vector having bits set only in those $w'_L$ tentative positions.
Finally, the overall $WF$ of SIA is obtained as
\begin{equation}
WF_{\mathrm{SIA}} = \sum_{i=1}^{w/w_L} WF_{\mathrm{SIA}}^{(i)}.
\label{eq:WFTSIA}
\end{equation}

When $w_L$ does not divide $w$, the only difference is in the $WF$ of the last step that, using the first one of the aforementioned strategies, can be obtained from eq. \eqref{eq:WFSIA}
by replacing $\binom{w - (i - 1)w_L}{w_L}$ with $1$. Using the second strategy, instead, the $WF$ of the last step must be computed as
\begin{multline}
WF_{\mathrm{SIA}}^{(\lceil w/w_L \rceil)} = \left\{ \left[ \frac{C_{s_L,1}}{P_\land(r, \left\{s_1, s_2, \ldots, s_L\right\}, w_L)} + C_{s_L,2} \right] \frac{\binom{r}{w_L}}{\binom{(\lceil w/w_L \rceil - 1)w_L}{w_L - w + (\lceil w/w_L \rceil - 1)w_L}} + C_{I \ge J,L} \right\} \\
\cdot \frac{1}{P_{I \ge J,L}(w_L, w_c,w'_L)}.
\label{eq:WFSIAb}
\end{multline}

Concerning the possible applicability of quantum speedups to forgery attacks based on support intersections, there is no apparent way of applying Grover's algorithm to speedup these attack procedures.
Therefore, it is reasonable to assume that no quantum speedups apply, i.e., $WF^{(pq)}_{\mathrm{SIA}} = WF^{(cl)}_{\mathrm{SIA}} = WF_{\mathrm{SIA}}$ as expressed by \eqref{eq:WFTSIA}.
However, in order to provide a conservative option, in the proposed instances of \sysacro{} we also consider the theoretical case in which the maximum possible speedup due to Grover's algorithm is achieved, i.e., $WF^{(pq)}_{\mathrm{SIA}} = \sqrt{WF^{(cl)}_{\mathrm{SIA}}} = \sqrt{WF_{\mathrm{SIA}}}$.

\subsection{Statistical attacks}
\label{sec:StatAttacks}

As shown in \cite{Phesso2016}, the sparse character of the matrix $\qcS$ and the statistical properties of the signature $\sigma$ expose the system to a statistical attack based on the collection of many signatures.
In fact, the signature is obtained as the sum of the columns of $\qcS$ (i.e., rows of $\qcS^T$) selected by the support of $(c+e)$.
Since both $\qcS$ and $(c+e)$ are sparse, the probability of cancellations between symbols 1 is very low.
This means that, if the signature bits in positions $i$ and $j$ are simultaneously set with relatively high probability, this is likely due to a column of $\qcS$ having two symbols 1 in these positions.
An opponent may then collect a sufficiently large number of signatures, in order to empirically compute covariances between bits.
Then, the dependence between the covariance value and the supports of the corresponding rows of $\qcS$ can be exploited to construct an alternative representation of the private key, which allows to forge sparse and valid signatures.

Let us denote as $Cov(\sigma_i,\sigma_j)$ the covariance between the $i$-th and $j$-th bits of the signature: if the support of a column of $\qcS$ contains both $i$ and $j$ (i.e., the supports of the $i$-th and $j$-th rows of $\qcS$ are not disjoint), then $Cov(\sigma_i,\sigma_j)$ will be larger with respect to the case in which the support of each column of $\qcS$ does not contain both $i$ and $j$ (i.e., the supports of the $i$-th and $j$-th rows of $\qcS$ are disjoint).
As in \cite{Phesso2016}, we neglect the case in which both $i$ and $j$ are included in the support of more than one column of $\qcS$, which is extremely rare due to the sparsity of $\qcS$.
A simple threshold criterion can then be used to separate the covariance values in two sets: if $Cov(\sigma_i,\sigma_j)$ exceeds some threshold $\Gamma$, then the opponent can conclude that there is a column in $\qcS$ having both $i$ and $j$ in its support.
This allows the construction of a graph $\mathcal{G}$ with $n$ nodes, having an edge between the $i$-th and $j$-th nodes only if $Cov(\sigma_i,\sigma_j) \geq \Gamma$.  
Once the graph has been obtained, the opponent can reconstruct a column-permuted version of $\qcS$ by exploiting cliques in the graph: indeed, each column of $\qcS$ identifies a set of $m_S$ pairwise connected nodes in the graph, forming a clique with size $m_S$.
This permuted version of $\qcS$ can be used to compute a permuted version of $\qcQ$, and by exploiting both of them the opponent can forge signatures which are sparser than authentic ones. Then a codeword of the public code must be added to achieve the same density of a valid signature but, as it is proven in \cite{Phesso2016}, this is not a serious issue.

This attack is obstructed by the presence of the random codewords, which influence the covariance distribution.
In fact, it might happen that $Cov(\sigma_i,\sigma_j)$ has a large value in some cases in which there is no column having $i$ and $j$ in its support, because of the interplay between the rows of $\qcG$ (used to generate the random codeword) and the columns of $\qcS$.
In fact, the signature can be written as
\begin{equation}
\label{eq:sig_comp}
\sigma=(c+e) \cdot \qcS^T=u\cdot \qcG\cdot \qcS^T + e\cdot \qcS^T=u \cdot \qcG_S +e\cdot \qcS^T.
\end{equation}
Thus, the signature is obtained as the sum of rows from $\qcG_S$ and rows from $\qcS^T$, so the covariance values also depend on the interplay between $\qcG$ and $\qcS$.
In order to take this into account, the covariance classification process can be performed by using two threshold values $\Gamma_1<\Gamma_2$. Then, two graphs are constructed:
\begin{itemize}
\item the graph $\mathcal{G}_1$, constructed using $\Gamma_1$, which must contain every edge belonging to $\mathcal{G}$;
\item the graph $\mathcal{G}_2$, constructed using $\Gamma_2$, which must not contain edges that are not in $\mathcal{G}$.
\end{itemize}
These graphs are constructed in such a way that 
\begin{equation}
\label{eq:GraphRel}
\mathcal{G}_2\subseteq\mathcal{G}\subseteq\mathcal{G}_1.
\end{equation}
The following rule is adopted to add edges to $\mathcal{G}_2$: an edge between $i$ and $k$ in $\mathcal{G}_2$ is added when there is a value of $j$ for which there are edges between $i$ and $j$ and
between $j$ and $k$ in $\mathcal{G}_2$ and $\left\{i; j; k \right\}$ forms a triangle in $\mathcal{G}_1$, meaning that there are edges between $i$ and $j$, between $i$ and $k$ and between $j$ and $k$ in $\mathcal{G}_1$.
With this procedure, the opponent adds edges to $\mathcal{G}_2$ until $\mathcal{G}_2 = \mathcal{G}$.
This attack procedure is based on the empirical estimate of the covariances between couples of bits, which can be obtained by the observation of several collected signatures.
In order to achieve sufficient statistical confidence of such estimates, the opponent needs to collect and analyze a sufficiently large number of signatures.
Next we describe an approach that allows computing a conservative and reliable lower bound on the number of needed signatures to perform the attack.

\subsubsection{Estimate of the attack efficiency}
\label{sec:lower_bound}
Let us define a lower bound on the number of signatures that an opponent needs to collect in order to make the attack successful.
In particular, we describe a method to compute the number of signatures for which the probability of success of the attack does not overcome a fixed threshold.
We first consider the case of codes with general structure, and then show how the \ac{QC} structure might yield a significant reduction in the attack complexity.
Since $S$ is sparse, we can assume that the maximum number of positions in which the supports of any two rows of $S^T$ (i.e., columns of $S$) intersect is equal to $1$. In such a case, each edge in $\mathcal{G}$ is associated only to one row of $S^T$; thus, since each row of $S^T$ corresponds to $\frac{m_S(m_S-1)}{2}$ edges, the total number of edges in $\mathcal{G}$ is equal to $n\frac{m_S(m_S-1)}{2}$.
As explained above, the opponent tries to guess $\mathcal{G}$ by adding edges to $\mathcal{G}_2$: if a row of $S^T$ is such that no edge is included in $\mathcal{G}_2$, then that row cannot be recovered through the attack.
We assume that, for the attack to be successful, all the rows of $S^T$ must be reconstructed by the opponent. 
If a row of $S^T$ is such that no edge is contained in $\mathcal{G}_2$, then the opponent has no information about that row.
He might try proceeding with an enumeration, but this procedure is infeasible, since each one of such rows corresponds to $\binom{n}{m_S}$ candidates. This means that the attack becomes feasible only when the rows of $S^T$ can be reconstructed through this graph procedure, and this depends on how $\mathcal{G}_2$ evolves.

If we denote as $\sigma_i$ and $\sigma_j$ the signature bits at positions $i$ and $j$, respectively, then the corresponding covariance is defined as
\begin{equation}
Cov(\sigma_i , \sigma_j) = E[\sigma_i\sigma_j]-E[\sigma_i]E[\sigma_j],
\label{eq:Covariance}
\end{equation}
where $E[\cdot ]$ denotes the mean value.
The opponent empirically computes covariances over the collected signatures. Let us denote as $\sigma^{(l)}$ the $l$-th collected signature, and as $\sigma_i^{(l)}$ and $\sigma_j^{(l)}$ its corresponding bits at positions $i$ and $j$.
Thus, the covariance in \eqref{eq:Covariance} can be rewritten as
\begin{align}
\label{eq:CovarianceBis}
Cov(\sigma_i,\sigma_j) & =\frac{1}{N}\sum_{l=1}^{N}{\sigma^{(l)}_i\sigma^{(l)}_j}-\frac{1}{N^2}\sum_{v=1}^{N}{\sigma^{(v)}_i}\sum_{u=1}^{N}{\sigma^{(u)}_j}  \\\nonumber
& = \frac{1}{N}X_{i,j}-\frac{1}{N^2}X_i X_j,
\end{align}
where $X_{i,j}$ is the number of signatures in which both bits at positions $i$ and $j$ are set, while $X_i$ (resp. $X_j$) is the number of signatures in which the bit at position $i$ (resp. $j$) is set.  Obviously, $X_{i,j} \leq X_i$ and $X_{i,j} \leq X_j$. 
We denote as $s_i$ and $s_j$, respectively, the $i$-th and $j$-th rows of $S^T$, and distinguish between the cases of $s_i$ and $s_j$ being disjoint or not.
Let us denote as $b$ the number of overlapping ones between these two rows. According to the above assumption, we have $b=0$ or $b=1$.
The pairs of indexes $(i,j)$ can be grouped in sets $\Im^{(b)}$, for $b=0,1$, that are defined as follows:
\begin{align}
\Im^{(1)} &= \left\{\left.(i,j)\in\mathbb{N}\times\mathbb{N}\right|\exists v<n \text{\hspace{1mm} s. t.\hspace{2mm}} i,j\in\phi(s_v)\right\}, \nonumber \\
\Im^{(0)} &= \left\{\left.(i,j)\in\mathbb{N}\times\mathbb{N}\right|\not\exists v<n \text{\hspace{1mm} s. t.\hspace{2mm}} i,j\in\phi(s_v)\right\},
\end{align}
where $\phi(\cdot)$ denotes the support of a vector.

Since the matrix $S$ is randomly generated, the values $X_i$ and $X_j$ can be described as the sum of Bernoulli random variables that are equal to $1$ with probability $\dot{\rho}$ and equal to $0$ with probability $1 - \dot{\rho}$.
Such a probability does not depend on the positions $i$ and $j$ and on the value of $b$, and can be computed as
\begin{equation}
\dot{\rho} = E[\sigma_i]=\sum^{m_S}_{l=1, \text{\hspace{1mm}$l$ odd}}{\frac{\binom{m_S}{l}\binom{n-m_S}{w'-l}}{\binom{n}{w'}}},
\end{equation}
where $w'$ denotes the weight of $c+e$, and so can be assumed equal to $(w+m_gw_g)$.

Analogously, $X_{i,j}$ can be described as the sum of $N$ Bernoulli random variables that are equal to $1$ with some probability that, however, depends on $b$. We denote such a probability as $\ddot{\rho}^{(b)}$.
In the case of $b = 1$, the event of a signature having the $i$-th and $j$-th bits simultaneously set can be due to two different phenomena: 
\begin{itemize}
\item the signature contains the $v$-th row of $S^T$, that is the one including both $i$ and $j$ in its support; we denote the probability of such an event as $\ddot{\rho}^{(1\wedge v)}$, with
\begin{equation}
\ddot{\rho}^{(1\wedge v)} = \frac{w'}{n}\sum_{\begin{smallmatrix}l = 0 \\\text{$l$ even}\end{smallmatrix}}^{m_S-1}\sum_{\begin{smallmatrix}u = 0 \\\text{$u$ even}\end{smallmatrix}}^{m_S-1}{\frac{\binom{m_S-1}{l}\binom{m_S-1}{u}\binom{n+1-2m_S}{w'-l-u-1}}{\binom{n-1}{w'-1}}};
\end{equation}
\item the signature does not contain the $v$-th row but still has the $i$-th and $j$-th bits that are simultaneously set; we denote the probability of such an event as $\ddot{\rho}^{(1\wedge\neg v)}$, with
\begin{equation}
\ddot{\rho}^{(1\wedge\neg v)} = \frac{n-w'}{n}\sum_{\begin{smallmatrix}l = 1 \\\text{$l$ odd}\end{smallmatrix}}^{m_S-2}\sum_{\begin{smallmatrix}u = 1 \\\text{$u$ odd}\end{smallmatrix}}^{m_S-2}{\frac{\binom{m_S-1}{l}\binom{m_S-1}{u}\binom{n+1-2m_S}{w'-l-u}}{\binom{n-1}{w'}}}.
\end{equation}
\end{itemize}
Then, we have
\begin{equation}
\ddot{\rho}^{(1)} = \ddot{\rho}^{(1\wedge v)} + \ddot{\rho}^{(1\wedge\neg v)}.
\end{equation}

For the case of $b=0$ we have
\begin{equation}
\ddot{\rho}^{(0)} = \sum_{\begin{smallmatrix}l = 1 \\\text{$l$ odd}\end{smallmatrix}}^{m_S}\sum_{\begin{smallmatrix}u = 1 \\\text{$u$ odd}\end{smallmatrix}}^{m_S}{\frac{\binom{m_S}{l}\binom{m_S}{u}\binom{n-2m_S}{w'-l-u}}{\binom{n}{w'}}}.
\end{equation}

Let us now take into account two couples of indexes $(i_1,j_1)\in\Im^{(1)}$ and $(i_0,j_0)\in\Im^{(0)}$, and consider the difference between the corresponding covariances computed according to \eqref{eq:CovarianceBis}, that is,
\begin{align}
&\frac{X_{i_1,j_1}}{N}-\frac{X_{i_1}X_{j_1}}{N^2}-\frac{X_{i_0,j_0}}{N}-\frac{X_{i_0}X_{j_0}}{N^2} = \nonumber \\
& \frac{X_{i_1,j_1}-X_{i_0,j_0}}{N}-\frac{X_{i_1}X_{j_1}-X_{i_0}X_{j_0}}{N^2}.
\end{align}
We can assume that, for sufficiently large values of $N$, we have $\frac{X_{i_1}X_{j_1}-X_{i_0}X_{j_0}}{N^2} \ll \frac{X_{i_1,j_1}-X_{i_0,j_0}}{N}$.
Indeed, we know that $E[X_{i_1}]=E[X_{i_0}]=E[X_{j_1}]=E[X_{j_0}]$: this means that the term $\frac{X_{i_1}X_{j_1}-X_{i_0}X_{j_0}}{N^2}$ is a random variable with null mean and variance which decreases as $N$ increases.
So, from now on we will neglect it: this means that, when comparing two covariance values, we can compare them just by looking at the values of $X_{i,j}$.
We will use this result in the following.

Let us now consider all the couples of indexes $(i,j)$ belonging to the same row of $S^T$, say the $v$-th one; these couples belong to $\Im^{(1)}$ and are associated to $\frac{m_S(m_S-1)}{2}$ covariances.
We define as $\rho_v$ the probability that at least one of the edges associated to such covariances is included in the graph $\mathcal{G}_2$.
First of all, we compute the probability that the maximum value of $X_{i,j}$ associated to these couples is equal to a integer $\bar{X}$.
In order to obtain this probability, we first need to compute other two probabilities:
\begin{itemize}
\item we define as $\rho_{X_{i,j}=\bar{X}}$ the probability that one $X_{i,j}$ is equal to $\bar{X}$. We can compute this probability as
\begin{equation}
\rho_{X_{i,j}=\bar{X}} = \binom{N}{\bar{X}}\left( \ddot{\rho}^{(1)}\right)^{\bar{X}}\left( 1 - \ddot{\rho}^{(1)}\right)^{N-\bar{X}};
\end{equation}
\item we define as $\rho_{X_{i,j}<\bar{X}}$ the probability that one $X_{i,j}$ is lower than $\bar{X}$. We can compute this probability as
\begin{equation}
\rho_{X_{i,j}<\bar{X}} = \sum_{l=0}^{\bar{X}-1}\binom{N}{l}\left( \ddot{\rho}^{(1)}\right)^{l}\left( 1 - \ddot{\rho}^{(1)}\right)^{N-l}.
\end{equation}
\end{itemize}
Then, the probability that the $X_{i,j}$ values associated to couples in the $v$-th row have maximum value equal to $\bar{X}$ can be computed as
\begin{equation}
\rho^*_{\bar{X}}=\sum_{l=0}^{\frac{m_S(m_S-1)}{2}}{\binom{\frac{m_S(m_S-1)}{2}}{l}\left( \rho_{X_{i,j}=\bar{X}}\right)^l \left( \rho_{X_{i,j}<\bar{X}}\right)^{\frac{m_S(m_S-1)}{2}-l}}.
\end{equation}
We conservatively suppose that the opponent is able to choose an optimal threshold for constructing the graph $\mathcal{G}_2$: this means that he will set such a threshold as the maximum value of the covariances associated to couples belonging to $\Im^{(0)}$.
Based on the above arguments on the values of $X_i$, we can just look at the values of $X_{i,j}$.
For a generic couple $(i,j)\in\Im^{(0)}$, the probability of the corresponding $X_{i,j}$ being lower than $\bar{X}$ is obtained as
\begin{equation}
\rho'_{\bar{X}} = \sum_{l=0}^{\bar{X}-1}{\binom{N}{l}\left(\ddot{p}^{(0)}\right)^l\left(1-\ddot{p}^{(0)}\right)^{N-l}}.
\end{equation}
Then, the probability that an edge from the $v$-th row of $S^T$ is included in $\mathcal{G}_2$ can be obtained as
\begin{equation}
\rho^{**}_{\bar{X}} = \left( \rho'_{\bar{X}} \right)^{\frac{n(n-1)}{2}-n\frac{m_S(m_S-1)}{2}}.
\end{equation}
Finally, we can compute the probability that at least one edge belonging to the $v$-th row of $S^T$ is put in $\mathcal{G}_2$ by considering all possible values of $\bar{X}$, that is:
\begin{equation}
\rho_{v} = \sum_{\bar{X}=0}^{N}{\rho_{\bar{X}}^*\rho_{\bar{X}}^{**}}.
\end{equation}

Thus, the probability that the graph $\mathcal{G}_2$ contains at least one edge for each row of $S^T$ can be obtained as $p_{\mathcal{G}_2} = \left( \rho_v\right)^n$.
This probability can be considered as a reliable lower bound on the probability of success of the attack.
Indeed, a row of $S^T$ can be reconstructed only when a sufficiently large number of its edges is put in $\mathcal{G}_2$. 
Estimating how much this number should be large to allow correct reconstruction might be quite hard.
However, it is clear that if no edges from a row are included in the graph, then that row cannot be reconstructed through the attack.

Let us now fix a probability equal to $2^{-\lambda}$, where $\lambda$ is the desired security level,  and compute the maximum number $N_{\lambda}$ of collected signatures for which the attack success probability does not overcome $2^{-\lambda}$. This number corresponds to
\begin{equation}
\label{eq:Nlambda}
N_{\lambda} = \max{\left\{N\text{\hspace{2mm}s.t.\hspace{2mm}} p_{\mathcal{G}_2}<2^{-\lambda} \right\}}.
\end{equation}

In the case of a matrix $\qcS$ with \ac{QC} structure, the opponent just needs to reconstruct a subset of all the rows of $\qcS^T$, since he can build the missing rows with cyclic shifts.
Indeed, the opponent just needs to determine one row in the first block of $p$ rows (i.e., the ones from $s_0$ to $s_{p-1}$), one row in the second block of $p$ rows (i.e., the ones from $s_p$ to $s_{2p-1}$).
Obviously, with respect to the non-\ac{QC} case, this leads to a speedup in the attack, meaning that the value of $N_{\lambda}$ gets reduced.
Let us consider a block of $p$ rows of $\qcS^T$: the probability that at least one row of such a block is included in $\mathcal{G}_2$ is
\begin{equation}
\tilde{p}_v = 1-\left( 1-p_v \right)^p.
\end{equation}
Thus, the probability that, in the \ac{QC} case, the graph $\mathcal{G}_2$ contains at least one edge for each row of $\qcS^T$ can be obtained as
\begin{equation}
\tilde{p}_{\mathcal{G}_2} = \left( \tilde{p}_v \right)^{n_0}.
\end{equation}
Similarly to \eqref{eq:Nlambda}, in this case the bound on the number of signatures is obtained as
\begin{equation}
\label{eq:qcNlambda}
\tilde{N}_{\lambda} = \max{\left\{N\text{\hspace{2mm}s.t.\hspace{2mm}} \tilde{p}_{\mathcal{G}_2}<2^{-\lambda} \right\}}.
\end{equation}

%
%
\section{System Instances}\label{sec:instances}
Based on the analysis of all the attacks studied in the previous section, 
we can design instances of \sysacro{} for achieving some fixed \ac{SL}.
As a guideline we use the five security categories, numbered from $1$ to $5$, defined within the NIST post-quantum standardization 
effort~\cite{NISTcall2016}.
Category $1$ corresponds to an \ac{SL} $\ge 2^{128}$, 
categories $2$--$3$ correspond to an \ac{SL} $\ge 2^{192}$ and categories $4$--$5$ correspond to an \ac{SL} $\ge 2^{256}$.
We point out that, given a target \ac{SL}, many secure system instances can be designed, having different parameters.
As a selection criterion, we first fixed a maximum value of the signature density $\delta_{\sigma}^{(\max)}$.
According to the signature generation of \sysacro{}, the signature density is equal to $\frac{m_S(w+m_gw_g)}{n}$.
So, we have chosen the system parameters $\left\{m_S, w, m_g, w_g, n\right\}$ in such a way as to achieve the largest possible signature density $\le \delta_{\sigma}^{(\max)}$ that allows achieving the target \ac{SL} of each security category against all the considered attacks.
In the following we propose nine instances of \sysacro{} we have selected, 
grouped in three classes corresponding to the three different security categories ,and characterized by a different value of the maximum signature density $\delta_{\sigma}^{(\max)}$.
Their parameters are reported in Table \ref{tab:sys_parameters}.

The instances named $a3$, $b3$, $c3$ and $a6$, $b6$, $c6$ have been designed taking into account quantum speedups of attacks, where applicable. 
As discussed in Sections \ref{LinearCombinationAttack} and \ref{sec:SIA}, 
for LCAs and SIAs no quantum speedups apparently exist.
Therefore, no quantum speedups have been considered for those two attacks.
Instead, the instances named $\alpha3$, $\beta3$, $\gamma3$ have been designed taking into account the maximum theoretical speedup which may follow from the application of Grover's algorithm to LCAs and SIAs.
Therefore, those instances of \sysacro{}, which consider $\delta_{\sigma}^{(\max)}=1/3$, are characterized by a complexity of LCAs and SIAs which is greater than $SL^2$.

The circulant size $p$ has been chosen as a prime number in order to avoid possible attacks based on the factorization of $p$, such as the one proposed in \cite{Shooshtari2016}. 
For the same reason, we have also chosen $r_0$ as a prime.
Increasing the value of $p$ allows a reduction in the public key size, but its 
value cannot exceed $r/(2w+1)$, otherwise the circulant blocks of matrix $V$ 
cannot all be circulant permutation matrices (since some block must have row 
weight greater than 1).
In addition, we have chosen values of $p$ as close as possible, but smaller 
than, a multiple of the assumed machine word size (i.e., $64$-bit), in order to minimize the amount of used machine word space in representing the $p \times p$ circulant blocks of any matrix as polynomials in $\mathbb{F}_2[x]/\langle x^p + 1\rangle$.
This has the advantage of allowing the data to fit better into the caches of the executing platform.

The values of $n_0$ have been chosen as primes such that the multiplicative order of $2$ modulo $n_0$ is $\mathtt{ord}_{n_0}(2)=n_0-1$, and $m_S$ is always odd; with these choices, we guarantee that $S$ is non singular, as explained in Section \ref{sec:KeyGen}.

\begin{table*}[!t]
\renewcommand{\arraystretch}{1.3}
\caption{Parameters of the proposed instances of \sysacro{}
\label{tab:sys_parameters}
}
\centering
\footnotesize
\begin{tabular}{ccc|ccccccccc}
\toprule
\textbf{Category} & $\mathbf{\delta_{\sigma}^{(max)}}$ & $\mathbf{ID}$ & $\mathbf{n_0}$     & $\mathbf{r_0}$ & $\mathbf{p}$  & $\mathbf{z}$   & $\mathbf{m_T}$ & $\mathbf{m_S}$ & $\mathbf{w}$ & $\mathbf{w_g}$ & $\mathbf{m_g}$ \\
 \midrule
\multirow{3}{*}{$1$} & $1/3$  & $a3$ & $227$ & $89$  & $127$  & $2$ & $1$  & $9$  & $42$ & $85$ & $11$ \\
	&$1/6$   & $a6$ & $139$ & $83$    &  $383$  & $2$ & $1$  & $9$  & $38$ & $77$ & $12$ \\
    &$1/3$   & $\alpha3$ & $149$ & $89$    &  $509$  & $2$ & $1$  & $23$  & $40$ & $81$ &  $13$ \\
    
\midrule                     
\multirow{3}{*}{$2$--$3$} & $1/3$   & $b3$ & $293$ & $149$  & $251$  & $2$ & $1$  & $13$  & $54$ & $109$ & $16$ \\
	&$1/6$   & $b6$ & $179$ & $113$  & $1279$  & $2$ & $1$  & $23$  & $46$ & $93$ & $17$ \\
    &$1/3$   & $\beta3$ & $173$ & $103$  & $1663$  & $2$ & $1$  & $43$  & $48$ & $97$ & $22$ \\
                        
\midrule                                           
\multirow{3}{*}{$4$--$5$} & $1/3$   & $c3$ & $269$ & $149$  & $571$  & $2$ & $1$  & $17$  & $72$ & $145$ & $20$ \\
	&$1/6$   & $c6$ & $211$ & $131$  & $3449$  & $2$ & $1$  & $43$  & $54$ & $109$ & $24$ \\
    &$1/3$   & $\gamma3$ & $293$ & $139$  & $3121$  & $2$ & $1$  & $69$  & $66$ & $133$ & $32$ \\
\bottomrule
\end{tabular}
\end{table*} 

In Table \ref{tab:sys_security} we provide the \ac{WF} of the proposed instances of \sysacro{}, computed on the basis of the attacks described in the previous sections.
For each attack, we consider attackers provided with both classical and quantum computers.
Any key pair of \sysacro{} must be renewed after generating a certain number of signatures to prevent statistical attacks of the type described in \cite{Phesso2016}.
A safe lifetime of any key pair can be estimated with the approach described in Section \ref{sec:lower_bound}.
In particular, if we want to provide a parameter set for a security level $SL$, then we can use eq. \eqref{eq:qcNlambda} and set $\lambda = SL$.
The corresponding lifetime is denoted as $\tilde{N}_{SL}$ in Table \ref{tab:sys_security}.

\begin{table*}[!t]
\renewcommand{\arraystretch}{1.3}
\caption{Number of different signatures, number of random codewords, attack work factors and key pair lifetime for the proposed instances of \sysacro{}
\label{tab:sys_security}
}
\centering
\footnotesize
\begin{tabular}{cc|ccccccccc}
\toprule
\textbf{Category} & $\mathbf{ID}$ & $\mathbf{N_s}$ & $\mathbf{A_{w_c}}$     & $\mathbf{WF_{SIA}}$ & $\mathbf{WF_{LCA}}$  & $\mathbf{WF_{DA}^{(pq)}}$   & $\mathbf{WF_{KRA}^{(pq)}}$ & $\mathbf{WF_{DA}^{(cl)}}$   & $\mathbf{WF_{KRA}^{(cl)}}$ & $\mathbf{\tilde{N}_{SL}}$\\

\midrule

\multirow{3}{*}{$1$} & $a3$  & $393.49$ & $129.81$ & $152.43$  & $209.87$  & $281.88$ & $540.18$  & $540.02$  & $>1000$ & $2655$\\
& $a6$  & $417.75$ & $143.82$ & $128.65$  & $227.56$  & $156.63$ & $276.93$  & $250.12$  & $499.23$ & $973$\\
& $\alpha3$  & $457.51$ & $161.14$ & $264.84$  & $259.39$  & $372.27$ & $719.39$  & $677.63$  & $>1000$ & $12002$\\

\midrule

\multirow{3}{*}{$2-3$} & $b3$  & $581.18$ & $198.01$ & $203.19$  & $308.49$  & $372.06$ & $715.38$  & $678.13$  & $>1000$ & $5571$\\
& $b6$  & $594.66$ & $229.87$ & $192.23$  & $348.86$  & $383.65$ & $732.48$  & $696.34$  & $>1000$ & $4851$\\
& $\beta 3$  & $629.57$ & $300.30$ & $394.86$  & $386.01$  & $805.70$ & $>1000$  & $>1000$  & $>1000$ & $34501$ \\

\midrule

\multirow{3}{*}{$4-5$} & $c3$  & $832.29$ & $260.20$ & $259.47$  & $433.79$  & $553.38$ & $>1000$  & $>1000$  & $>1000$ & $8790$\\
& $c6$  & $775.34$ & $354.73$ & $266.47$  & $474.62$  & $833.40$ & $>1000$  & $>1000$  & $>1000$ & $14269$\\
& $\gamma 3$  & $925.90$ & $486.32$ & $517.65$  & $587.59$  & $>1000$ & $>1000$  & $>1000$  & $>1000$ & $107005$\\

\bottomrule
\end{tabular}
\end{table*} 

%
%
\section{Benchmarks on a NIST Compliant Platform}\label{sec:performance}
\begin{table*}[!t]
\small
\begin{center}
\caption{Running times for key generation, signature and verification as a 
function of the chosen category density (either $\frac{1}{3}$ or $\frac{1}{6}$) 
and Grover speedups (denoted by an identifier expressed with a greek lowercase 
letter) on an AMD Ryzen 5 1600 CPU at 3.2 GHz.\label{tab:runningtimes}}
\begin{tabular}{cc|cccc}
\toprule
\multirow{2}{*}{\textbf{Category}} & \multirow{2} {*} {\textbf{ID}}  &  \textbf{KeyGen}  & \textbf{Sign}   & \textbf{Sign+Decompress}  & \textbf{Verify} \\
                                   &                                 &  \textbf{(ms)}    &  \textbf{(ms)}  &  \textbf{(ms)}            & \textbf{(ms)}   \\
\midrule
\multirow{3}{*}{1}        &  $a_3$     &    35.51   ($\pm$ 0.96)  & 0.29   ($\pm$ 0.02) & 1.96   ($\pm$ 0.06)  & 28.71   ($\pm$ 1.12)  \\
                          & $a_6$      &    27.23   ($\pm$ 1.18)  & 0.14   ($\pm$ 0.01) & 1.06   ($\pm$ 0.04)  & 31.18   ($\pm$ 0.56)  \\ 
                          & $\alpha_3$ &    43.45   ($\pm$ 2.24)  & 0.28   ($\pm$ 0.04) & 1.52   ($\pm$ 0.04)  & 51.10   ($\pm$ 1.51)  \\
\midrule
\multirow{3}{*}{2--3}     & $b_3$      &    154.49  ($\pm$ 5.45)  & 0.27   ($\pm$ 0.02) & 2.29   ($\pm$ 0.08)  & 97.83   ($\pm$ 1.93)  \\
                          & $b_6$      &    227.14  ($\pm$ 10.67) & 0.55   ($\pm$ 0.03) & 2.30   ($\pm$ 0.13)  & 179.89  ($\pm$ 2.68)  \\
                          & $\beta_3$  &    249.69  ($\pm$ 12.04) & 1.11   ($\pm$ 0.06) & 2.62   ($\pm$ 0.11)  & 212.19  ($\pm$ 1.66)  \\
\midrule
\multirow{3}{*}{4--5}     & $c_3$      &    290.95  ($\pm$ 11.68) & 0.71   ($\pm$ 0.04) & 5.97   ($\pm$ 0.19)  & 186.30  ($\pm$ 1.63)  \\
                          & $c_6$      &    840.74  ($\pm$ 31.04) & 2.59   ($\pm$ 0.09) & 3.81   ($\pm$ 0.12)  & 650.78  ($\pm$ 5.31)  \\
                          & $\gamma_3$ &   1714.01  ($\pm$ 69.22) & 4.27   ($\pm$ 0.14) & 9.16   ($\pm$ 0.20)  & 926.35  ($\pm$ 5.70)  \\
\bottomrule
 \end{tabular}
 \end{center}
\end{table*}

For the sake of completeness, we provide the results of a set of execution time 
benchmarks performed on the reference implementation that is publicly available 
in \cite{LEDAcrypt}.
Since no platform specific optimizations are exploited, we expect these results 
to be quite consistent across different platforms.

The results were obtained measuring the required time for key generation, 
signature and verification as a function of the chosen security category, 
the density of the signature ($\frac{1}{3}$ or $\frac{1}{6}$), and whether
or not Grover speedups can be attained on the SIA. 
The measurements reported are obtained as the average of $100$ executions of the 
reference implementation compiled with \texttt{gcc} 6.3.0 from Debian $9$ amd64.
Given the NIST requirement on the reference computing platform (an Intel 
x86\_64 CPU) we instructed \texttt{gcc} to employ the most basic instruction
set among the ones fitting the description (\texttt{-march=nocona} option). 
The generated binaries were run on an AMD Ryzen 5 1600 CPU at 3.2 GHz, locking the 
frequency scaling to the top frequency.
Table~\ref{tab:runningtimes} reports the obtained running times and standard 
deviations over the measured $100$ executions.
The main bottleneck of the computation for the key generation primitive is the 
materialization of the public code matrix $H'$, as its size exceeds the first 
level of caches. A similar issue is also present in the signature verification
primitive, where such a matrix is employed.
By contrast, the signature primitive is significantly faster, and sub-millisecond
execution times for all the instances belonging to category $1$.
We also report in Table~\ref{tab:runningtimes} the running times of the signature
primitive in case compressed private keys are employed. In this case, only
the value of the seed generated by a True Random Number Generator (TRNG) and the 
value of $B^T$ are stored as the private key of the cryptosystem, while all the 
other values are computed again before the signature takes place. Although this 
approach raises the amount of time required by the signature primitive, the total 
running time does not exceed $10$ ms even in the case of the category $5$ 
instance assuming that it is possible to apply Grover's algorithm to SIAs.

\begin{table*}[!t]
\small
\begin{center}
\caption{Sizes of the key pair,as a 
function of the chosen category density (either $\frac{1}{3}$ or $\frac{1}{6}$) 
and Grover speedups (denoted by an identifier expressed with a greek lowercase 
letter). We report both the compressed private key composed of the PRNG seed 
and $B^T$ only, and the entire expanded private key, and the signature as a 
function of the chosen category, density and Grover speedups.\label{tab:keysizes}}
 \begin{tabular}{cc|cccc}
 \toprule
\multirow{2}{*}{\textbf{Category}} & \multirow{2} {*} {\textbf{ID}}  & \multicolumn{2}{c}{\underline{\textbf{Private Key Size }}} & \textbf{Public Key}  & \textbf{Signature} \\
                       &                                          &  \textbf{At rest}   &  \textbf{In memory } &  \textbf{size}   & \textbf{size}  \\
                       &                                          &  \textbf{(B)}       &  \textbf{(kiB)}      &  \textbf{(kiB)}  & \textbf{(kiB)}  \\
                       \midrule                             
\multirow{3}{*}{1}    & $a_3$      & 56 &  53.66 &     315.67 & 3.55  \\ 
                      & $a_6$      & 56 &  21.89 &     540.80 & 6.52  \\ 
                      & $\alpha_3$ & 56 &  32.54 &     828.81 & 9.32  \\ 
\midrule                                   
\multirow{3}{*}{2--3} & $b_3$      & 64 &  76.29 &    1364.28 & 9.16  \\ 
                      & $b_6$      & 80 &  40.30 &    3160.47 & 27.98 \\ 
                      & $\beta_3$  & 64 &  55.77 &    3619.48 & 35.15 \\ 
\midrule                                   
\multirow{3}{*}{4--5} & $c_3$      & 88 &  86.03 &    2818.20 & 18.92  \\ 
                      & $c_6$      & 88 &  69.79 &   11661.05 & 89.02  \\ 
                      & $\gamma_3$ & 88 & 159.01 &   15590.80 & 112.17 \\
 \bottomrule
 \end{tabular}
\end{center}
\end{table*}

Finally, in Table~\ref{tab:keysizes} we report the sizes of the key pairs for 
all the proposed instances of \sysacro{}, together with the sizes of the 
corresponding signatures. We note that the compressed private keys, denoted in 
the table as \emph{at rest} are significantly smaller than their expanded 
counterpart, and can be easily fit even in constrained devices. 
%
%
\section{Concluding Remarks}\label{sec:conclusion}
We have designed and implemented an efficient post-quantum digital signature 
scheme based on low-density generator matrix codes.
The complexity of all known attacks against this system has been estimated 
taking into account both classical and quantum computers.
Recent statistical attacks have also been considered in the estimate of the 
lifetime of a key pair.
This has allowed designing several instances of the system which achieve given 
security levels.
Efficient algorithmic solutions have been proposed for the main functions of 
this system, which have been used in a software implementation that has been 
made publicly available.
Performance benchmarks run on this reference implementation show that the proposed
signature scheme can be efficiently implemented on general purpose computing 
platforms and achieves compact key pairs.
%
%
\appendix
%
%
\section{Alternative Expression for \texorpdfstring{$\qcQ^{-1}$}{inverse quasi cyclic Q}}
\label{app:qcQ}
Let us consider the matrices defined in Section \ref{subsubsec:qcQ}, and let us 
focus on the product between $A^*$ and $D^{-1}$. We have
\begin{align}
A^*\cdot D^{-1}&=\left(A\otimes 1_{p\times 1}\right)\cdot D^{-1}\nonumber \\
&=\left(A\cdot D^{-1}\right)\otimes 1_{p\times 1}.
\end{align}   
Multiplying by $B^{*T}$, we obtain
\begin{align}
A^*\cdot D^{-1}\cdot B^{*T}&=\left[\left(A\cdot D^{-1}\right)\otimes 1_{p\times 1}\right]\cdot \left(B^T\otimes 1_{1\times p}\right)\nonumber \\
&=\left(A\cdot D^{-1}\cdot B^T\right)\otimes\left(1_{p\times 1}\cdot 1_{1\times p}\right)\nonumber \\
&=\left(A\cdot D^{-1}\cdot B^T\right)\otimes 1_{p\times p}.
\end{align}
Let us denote $\left(A\cdot D^{-1}\cdot B^T\right)\otimes 1_{p\times p}$ as $\left(A\cdot D^{-1}\cdot B^T\right)^*$.
Equation \eqref{eq:Qinv} can then be rewritten as
\begin{equation}
\qcQ^{-1}=\qcM^T+\qcM^T\left(A\cdot D^{-1}\cdot B^T\right)^*\qcM^T.
\end{equation}
Considering the fact that $1_{p\times p}$ is invariant to permutations, we can write
\begin{equation}
\label{eq:Qinv_def}
\qcQ^{-1}=\qcM^T+\left(\Pi^T\cdot A\cdot D^{-1}\cdot B^T\cdot\Pi^T\right)\otimes 1_{p\times p}.
\end{equation}
When $p$ is even, we have $D = I_z$, and so we can further simplify eq. \eqref{eq:Qinv_def}, obtaining
\begin{align}
\qcQ^{-1}&=\qcM^T+\left(\Pi^T\cdot A\cdot I_z\cdot B^T\cdot\Pi^T\right)\otimes 1_{p\times p} \nonumber \\
&=\qcM^T+\left(\Pi^T\cdot A \cdot B^T\cdot\Pi^T\right)\otimes 1_{p\times p}.
\end{align}
%
%
\section{Probability of Success of a Single Step of a SIA}\label{app:PSIA}
In order to justify expression \eqref{eq:PSIA} for $P_{\mathrm{SIA}}$, let us focus on the probability of success of a single step of a SIA, that is, the probability that an attacker finds a vector $s_L$ with weight $w_L$ and the $w_L'$ bits in $e'$ that correspond to the $w_L$ bits set in $s_L$. 
For this purpose, we need to introduce some notation and definitions.

Given two binary vectors, $v_1$ and $v_2$, with length $n$ and weight $w_1$ and $w_2$,
respectively, the probability that their bit-wise AND, $v_1 \land v_2$, has weight $x$ is
\begin{equation}
P_\land(n, \left\{v_1, v_2\right\}, x) = \frac{\binom{w_1}{x}\binom{n-w_1}{w_2-x} }{ \binom{n}{w_2}}.
\label{eq:Pa2vectors}
\end{equation}
In this case, the bit-wise XOR of $v_1$ and $v_2$, $v_1 \oplus v_2$, has weight $y = w_1 + w_2 - 2x$.
Hence, in general, the probability that their bit-wise XOR has weight $y$ is
\begin{equation}
P_\oplus(n, \left\{v_1, v_2\right\}, y) = \frac{  \binom{w_1 }{ \frac{w_1 + w_2 - y}{2}} \binom{n-w_1}{w_2-\frac{w_1 + w_2 - y}{2}} }{ \binom{n}{w_2}}
\label{eq:Px2vectors}
\end{equation}
where $y$ must be such that $w_1 + w_2 - y$ is even.

In order to extend eq. \eqref{eq:Pa2vectors} and eq. \eqref{eq:Px2vectors} to
the case of more than two vectors, we can proceed as follows.
Let us consider the vector $v_x = v_1 \land v_2$, having weight
$x \in \left\{0, 1, 2, \ldots, \min\left(w_1, w_2\right) \right\}$ with
probability $P_\land(n, \left\{v_1, v_2\right\}, x)$.
If we add a third vector $v_3$ with weight $w_3$, and compute $v_1 \land v_2 \land v_3$,
the resulting vector can have weight $y \in \left\{0, 1, 2, \ldots, \min\left(w_1, w_2, w_3\right) \right\}$.
The probability that this occurs can hence be computed as
\begin{equation}
P_\land(n, \left\{v_1, v_2, v_3\right\}, y) = \sum_{x = y}^{\min\left(w_1, w_2\right)} P_\land(n, \left\{v_1, v_2\right\}, x) P_\land(n, \left\{v_x, v_3\right\}, y). 
\label{eq:Pa3vectors}
\end{equation}
Similarly, for the bit-wise XOR of three vectors, we have
\begin{equation}
P_\oplus(n, \left\{v_1, v_2, v_3\right\}, y) = \sum_{x = \left| w_1 - w_2 \right|}^{\min\left(w_1+w_2, n\right)} P_\oplus(n, \left\{v_1, v_2\right\}, x) \cdot P_\oplus(n, \left\{v_x, v_3\right\}, y),
\label{eq:Px3vectors}
\end{equation}
where $v_x$ denotes a weight-$x$ vector obtained as $v_1 \oplus v_2$
and $\left| \cdot \right|$ returns the absolute value of its argument.
The iterated application of eq. \eqref{eq:Pa3vectors} and eq. \eqref{eq:Px3vectors}
allows to compute these probabilities for an arbitrary number of
vectors.

By using these tools, we can estimate the probability of success of a SIA as follows.
The probability that $s_L$ has weight $w_L \ge d$ can be computed
through the iterated application of \eqref{eq:Pa3vectors}.
The iterated application of \eqref{eq:Px3vectors}, instead, allows to compute
the average weight of $s' = T \cdot s$.

Let us suppose that $s_L$ has weight $w_L \ge d$, and
let us denote by $I$ the set of $w'_L$ bits in $e'$ 
that correspond to the $w_L$ bits set in $s_L$.
For small values of $w_L$, like those necessary to minimize the
\ac{WF}, it can be considered that $I$ contains a
number of bit positions equal to $w'_L \approx m_s w_L$.
The vector $e'$ is obtained by summing the rows of $S^T$
(each with weight $m_s$) that are selected by the bits set within $e$.

The weight of $e$ is the same as that of $s'$, that is, $w$
in our case.
We focus on one of the $w_L \le w$ bits set in $e$ which
correspond to the intersected syndrome vector $s_L$ and aim at
estimating the probability that any of its corresponding $m_s$
bits is set in any of the intercepted vectors $e'$.
The position of the bit under exam gives the index of one row of $S^T$
that is included in the sum, and we focus on this row.
It contains $m_s$ bits, and we now aim at estimating the probability
that any of them is set in the vector resulting from the sum with the
other $(w - w_L)$ rows of $S^T$.
This occurs if and only if, at its corresponding column index,
the $(w - w_L)$ other selected rows of $S^T$
contain an even number of ones.
This way, we are considering that the supports of the considered
$w_L$ rows of $S^T$ are disjoint, that is, they have
no overlapping bits set.
So, we neglect a further mechanism that could cause cancellation
of some of the bits that are of interest for an attacker, thus we
are considering a facilitated scenario for him.
We also consider the case $w_L = d$, which provides the smallest
values of the \ac{WF}.

Based on these considerations, and taking into account that $e$
can have its $w$ ones only within the last $r$ bits, the probability
that any of the $m_s$ bits of interest is set in the vector resulting from 
the sum of $w$ rows of $S^T$ is
\begin{equation}
P_{I_1}(w_L)=\sum_{i=0,\, i\, \mathrm{even}}^{\min(l-1,(w - w_L))}\frac{ \binom{l-1}{i} 
\binom{r - w_L - l + 1}{(w - w_L) - i} }{ \binom{r - w_L }{ (w - w_L)}},
\label{eq:PI1}
\end{equation}
where we have considered that the last block of $r$ rows
in $S^T$ has column weight $l = \lfloor m_s r/n + 0.5 \rfloor$.

Then we must consider the addition of the random codeword $c$.
For this purpose, let us define the following probability
\begin{equation}
P_{I_2}(x,y)=\sum_{i=0,\, i\, \mathrm{even}}^{\min(x,y)}\frac{ \binom{y}{i} 
\binom{n - 1 - y }{ x - i} }{ \binom{n-1}{x}}.
\label{eq:PI2}
\end{equation}
Let us suppose that the support of $c$ does not include the position 
of the bit set in $e$ that is under exam.
In this case, the addition of $c$ has no effect on the considered bit
in $e'$ if the support of $c$ selects a set of $w_c$ rows of $S^T$
having an even number of ones in the column corresponding to that bit.
This occurs with probability $P_{I_2}(w_c,m_s-1)$.
Dually, any bit $\in I$ may be canceled in the sum of the $w$ rows of $S^T$
selected by the bits set in $e$ (this occurs with probability $1 - P_{I_1}(w_L)$)
and reappear in the sum of the $w_c$ rows of $S^T$ selected by $c$
(this occurs with probability $1 - P_{I_2}(w_c,m_s-1)$).
A similar reasoning can be applied when the support of $c$ includes the position 
of the bit set in $e$ that is under exam.
In this case, the probability that the other bits of the support of $c$
select a set of $w_c-1$ rows of $S^T$ having an even number of ones
in the column of interest is $P_{I_2}(w_c-1,m_s-1)$.

Based on these arguments, we can estimate the probability that any of the
$w'_L$ bits $\in I$ is set in $e'$ as follows
\begin{multline}
P_{I}(w_L, w_c) = \left\{P_{I_1}(w_L)P_{I_2}(w_c,m_s-1) + \left[1 - P_{I_1}(w_L)\right]\left[1 - P_{I_2}(w_c,m_s-1)\right] \right\} \left(1 - \frac{w_c}{n}\right) \\
+ \left\{P_{I_1}(w_L)\left[1 - P_{I_2}(w_c-1,m_s-1)\right] + \left[1 - P_{I_1}(w_L)\right]P_{I_2}(w_c-1,m_s-1) \right\} \frac{w_c}{n},
\label{eq:PIz}
\end{multline}
where $\left(1 - \frac{w_c}{n}\right)$ is the probability that the support of $c$
does not include the position of the bit set in $e$ that is under exam.

We denote by $J$ the complement set of $I$, and aim at estimating the probability
that, by counting the number of occurrences of each bit set within the ensemble of
intercepted signatures $e'$, an attacker is able to distinguish the bits $\in I$ from those $\in J$.
For this purpose, we can reason as before, with the main difference that we
do not need to focus on any specific bit of $e$.
This way, we obtain that the probability that there is a bit set in $e'$ at a
position $\in J$ is
\begin{equation}
P_{J}(w_L, w_c) = P_{J_1}(w_L)P_{I_2}(w_c,m_s) + \left[1 - P_{J_1}(w_L)\right] \left[1 - P_{I_2}(w_c,m_s)\right],
\label{eq:PJz}
\end{equation}
where
\begin{equation}
P_{J_1}(w_L)=\sum_{i=1,\, i\, \mathrm{odd}}^{\min(l,(w-w_L))
}\frac{\binom{l }{i}
\binom{r - w_L - l}{ (w - w_L) - i}}{ \binom{r - w_L}{(w - w_L)}}.
\label{eq:PJ1}
\end{equation}

Starting from eq. \eqref{eq:PIz} and eq. \eqref{eq:PJz}, we can estimate the
probability that a bit in $I$ or $J$ is set exactly $x$ times within
a group of $L$ vectors $e'$
\begin{equation}
P_{X,L,x}(w_L, w_c)  =  \binom{L}{x} P_{X}(w_L, w_c)^{x} \cdot \left[1-P_{X}(w_L, w_c)\right]^{L-x},
\label{eq:PIJL}
\end{equation}
where $X=I$ or $X=J$.

Based on eq. \eqref{eq:PIJL}, we can compute the probability that all the bits
at positions $\in I$ are set at least $x$ times within the $L$ vectors 
$e'$ (at least one is set exactly $x$ times)
\begin{equation}
P_{I,L,\ge x}(w_L, w_c,w'_L) = \left(\sum_{i=x}^{L}P_{I,L,i}(w_L, w_c)\right)^{w'_L} - \left(\sum_{i=x+1}^{L}P_{I,L,i}(w_L, w_c)\right)^{w'_L}. 
\label{eq:PILgex}
\end{equation}
Similarly, we can compute the probability that all bits at positions
$\in J$ are set at most $x$ times within the $L$ vectors $e'$
\begin{equation}
\label{eq:PJLlex}
P_{J,L,\le x}(w_L, w_c,w'_L) = \left(\sum_{i=0}^{x}P_{J,L,i}(w_L, w_c)\right)^{n-w'_L}.
\end{equation}
Based on these formulas, we can compute the probability that,
within the $L$ vectors $e'$, all the bits at positions $\in I$ are set
more times than the bits at positions $\in J$
\begin{equation}
P_{I \ge J,L}(w_L, w_c,w'_L) = \sum_{i=0}^{L-1} P_{J,L,\le i}(w_L, w_c,w'_L) \cdot P_{I,L,\ge i+1}(w_L, w_c,w'_L). 
\label{eq:PIgeJ}
\end{equation}

By multiplying eq. \eqref{eq:PIgeJ} by $P_\land(r, \left\{s_1, s_2, \ldots, s_L\right\}, w_L)$, computed through the iterated application of eq. \eqref{eq:Pa3vectors},
we obtain $P_{\mathrm{SIA}}$ as expressed by eq. \eqref{eq:PSIA}.
%
%
\bibliographystyle{unsrt}
\bibliography{Archive}
%
%
\end{document}